\documentclass[aps,twocolumn,groupedaddress]{revtex4-2}
\usepackage{graphicx}
\usepackage{amsmath}
\usepackage{amssymb}
\usepackage{braket}
\usepackage{import}
\usepackage{siunitx}
\usepackage{color}
\usepackage[normalem]{ulem}
\usepackage{multirow}
\usepackage{booktabs}
\usepackage{tikz}
\usetikzlibrary{calc,patterns,angles,quotes}
\usepackage{tikz-3dplot}
\usepackage{pgfplots,pgfplotstable}
\usepackage{comment}
\usepackage{hyperref}
\hypersetup{
	colorlinks=true,
	linkcolor=blue,
	filecolor=magenta,      
	urlcolor=blue,
	citecolor=blue,
}
\usepackage{cleveref}
\crefname{equation}{Eq.}{Eqs.}  
\Crefname{equation}{Equation}{Equations}	
\crefname{figure}{Fig.}{Figs.}
\Crefname{figure}{Figure}{Figures}
\crefname{chapter}{Ch.}{Chs.}
\Crefname{chapter}{Chapter}{Chapters}
\crefname{section}{Sec.}{Secs.}
\Crefname{section}{Section}{Sections}
\crefname{appendix}{App.}{App.}
\Crefname{appendix}{Appendix}{Appendices}	
\crefname{algorithm}{Alg.}{Algs.}
\Crefname{algorithm}{Algorithm}{Algorithm}
\crefname{table}{Table}{Tables}
\Crefname{table}{Table}{Tables}

\let\originalleft\left
\let\originalright\right
\renewcommand{\left}{\mathopen{}\mathclose\bgroup\originalleft}
\renewcommand{\right}{\aftergroup\egroup\originalright}

\begin{document}
	\cmidrulewidth=.03em
	
\title{Efficient ionization of two dimensional excitons by intense single cycle terahertz pulses}

\author{H{\o}gni C. Kamban}
\email{hck@mp.aau.dk}
\affiliation{Department of Materials and Production, Aalborg University, DK-9220 Aalborg \O st, Denmark}
\affiliation{Center for Nanostructured Graphene (CNG), DK-9220 Aalborg \O st, Denmark}
\author{Thomas G. Pedersen}
\affiliation{Department of Materials and Production, Aalborg University, DK-9220 Aalborg \O st, Denmark}
\affiliation{Center for Nanostructured Graphene (CNG), DK-9220 Aalborg \O st, Denmark}

\date{\today}

	\begin{abstract}
		External electric fields are highly attractive for dynamical manipulation of excitons in two-dimensional materials. Here, we theoretically study the ionization of excitons in monolayer transition metal dichalcogenides (TMDs) by intense pulsed electric fields in the terahertz (THz) regime. We find that THz pulses with realistic field strengths are capable of ionizing a significant fraction of photogenerated excitons in TMDs into free charge carriers. Short THz pulses are therefore an efficient, non-invasive method of dynamically controlling the free carrier concentration in monolayer TMDs, which is useful for applications such as THz modulators. We further demonstrate that exciton ionization probabilities should be experimentally measurable by comparing free carrier absorption before and after the THz pulse. Detailed results are provided for different TMDs in various dielectric environments.
	\end{abstract}

\maketitle

Monolayer transition metal dichalcogenides (TMDs) are exciting materials for optoelectronic applications \cite{Schaibley2016,Gong2017}. They are promising components in applications such as photodetectors \cite{Yin2012a,Lopez2013,Wang2015a}, THz modulators \cite{Cao2016,Fan2020}, and solar cells \cite{Bernardi2013,Lopez-Sanchez2014}, where they can absorb up to 5-10\% of incident sunlight in a thickness less than 1 nm \cite{Bernardi2013}. One of the most important characteristics of monolayer TMDs is the strongly bound excitons that form due to the reduced screening in two-dimensional (2D) materials \cite{Ramasubramaniam2012,Berkelbach2013,Olsen2016}. The optical properties of 2D materials are completely dominated by these excitons \cite{Wang2012b,Ramasubramaniam2012,Qiu2013,Trolle2014}, and methods of manipulating them are therefore highly sought after. These methods may be as simple as altering the structural design of the device components by controlling, e.g., substrate screening. However, a major disadvantage with these approaches is that the properties are fixed once the components have been constructed. A more attractive option is therefore to control the properties using external fields that may be switched on or off at will, and thereby obtain dynamic control of the material properties.

In recent years, interest in applying static in-plane electric fields to excitons in TMD monolayers, multilayers, and van der Waals heterostructures has been increasing \cite{Haastrup2016,Scharf2016a,Massicotte2018,Klots2014,Kamban2019,Pedersen2016b,Kamban2020a}. This processes induces electrons and holes that form the excitons to ionize into free charge carriers, and the result is a measurable increase in photocurrent generation \cite{Massicotte2018}. However, to apply the electric field, the authors of Ref. \cite{Massicotte2018} incorporated buried electrical contacts with a tiny gap into their TMD sample. This is by no means a trivial task, as it involves making complicated modifications to the sample. Consequently, static electric fields become less attractive as a means of manipulating excitons. 


With the rapid progression of THz technology, it is natural to ask how efficient pulsed THz electric fields are at inducing exciton ionization. A significant advantage of using THz pulses rather than static fields for this purpose is that no modifications need to be made to the sample. Whereas THz induced ionization of atoms in gaseous samples is typically measured by counting the number of ions produced during the pulse \cite{Jones1993,Li2014b}, the solid-state equivalent is typically discussed in electroabsorption experiments \cite{Ewers2012,Stein2018a,Murotani2018}. In these experiments, the absorption spectrum of a sample is measured in the presence of an electric field. In this case, one finds a shift and broadening of the exciton absorption peaks that depend on the strength of the electric field. The shift is well explained by the exciton Stark effect \cite{Klein2016,Pedersen2016a,Scharf2016a,Cavalcante2018}, while the broadening is commonly attributed to the reduced exciton lifetime due to field-induced exciton ionization \cite{Dow1970,Miller1985,Stein2018a}. However, detailed interpretations of recent electroabsorption experiments on both monolayer MoS$_2$ \cite{Shi2020} and carbon nanotubes \cite{Ogawa2010} reveal that exciton ionization is not the dominating contribution to this broadening. The authors of Refs. \cite{Ogawa2010,Shi2020} base their arguments on the fact that the measured broadening is proportional to the square of the electric field strength, which is not predicted by exciton ionization \cite{Massicotte2018,Kamban2019}. The apparent contradiction is resolved by noting that the contribution to the broadening by exciton ionization alone \cite{Haastrup2016,Massicotte2018,Kamban2019} is much lower than the field-induced phonon contribution \cite{Shi2020}. As a result, it is very difficult to measure exciton ionization in electroabsorption experiments. 

In the present paper, we theoretically study exciton ionization in two-dimensional TMDs induced by intense single-cycle THz pulses. We find that nearly all ionization occurs within a very short time interval near the peak field strength of the THz pulse. For the longest pulse duration considered here, this ionization interval is about $0.5$ ps, which is much shorter than the typical field-free exciton lifetimes of a few to hundreds of picoseconds in the popular TMDs WS$_2$ \cite{Yuan2015}, WSe$_2$ \cite{Cadiz2018,Massicotte2018,Mouri2014}, and MoS$_2$ \cite{Shi2013a,Korn2011}. We further show that a THz pulse with realistic field strength is capable of transforming a considerable portion of photogenerated excitons into free charge carriers in a realistic monolayer TMD. This process happens over just a few picoseconds, and therefore suggests that THz pulses are incredibly efficient for obtaining dynamic control over the number of free charge carriers, which is highly relevant for applications such as THz modulators \cite{Cao2016,Fan2020}. Additionally, such a substantial change in the number of free charge carriers should be measurable in the free carrier absorption of a sample, providing a direct method of estimating the exciton ionization probabilities in experiments. Our analysis clearly demonstrates that THz ionization of excitons in two-dimensional materials is feasible and we provide quantitative estimates of the yield. To the best of our knowledge, previous works do not consider these phenomena and we hope our results will inspire attempts at experimental verification.

\section{Simulation details}

Throughout the present paper, excitons will be modeled as electron-hole pairs that are screened by the TMD sheet as well as the surrounding dielectrics. As is commonly done, we shall approximate the dielectric function by the linearized form $\varepsilon\left(\boldsymbol{q}\right) = \kappa + r_0q$, where $\boldsymbol{q}$ is the momentum space coordinate, $\kappa$ the average dielectric constant of the surrounding dielectrics, and the screening length $r_0$ is related to the two-dimensional polarizability of the TMD monolayer by $r_0 = 2\pi\alpha_{\mathrm{2D}}$ \cite{Cudazzo2011a,Trolle2017}. Under this approximation, the electron-hole interaction potential is given by the Rytova-Keldysh form \cite{Rytova1967,Keldysh1979,Trolle2017,Kamban2020b}. By further making the two-band, effective mass approximation, the excitons may be described by the two-dimensional Wannier equation \cite{Wannier1937,Lederman1976a}, which has been shown to accurately reproduce the exciton binding energies computed by more numerically demanding methods in various 2D materials \cite{Cudazzo2010,Pulci2012a,Latini2015}. We use the experimentally verified material parameters from Ref. \cite{Goryca2019} throughout.

To study exciton ionization induced by THz pulses, a time-dependent field term is included in the Wannier equation. We have obtained a trace of the experimental THz pulse used in Ref. \cite{Shi2020}, and the shape of the pulse used throughout the present paper is based on this experimental pulse. For performing computations, we assume that only the excitonic ground state is occupied initially, and then let the wave function $\psi\left(t\right)$ evolve in time. The ionization probability $P_{\mathrm{ion}}$ is computed by subtracting the occupation probabilities of the bound states $P_b$ from unity, i.e. $P_{\mathrm{ion}}\left(t\right) = 1- \sum_b P_b\left(t\right)$. The bound states are the states with energy less than zero, and their occupation probabilities are given by $P_b=\left|\braket{\varphi_b|\psi\left(t\right)}\right|^2$. We make numerical computations feasible by forcing the functions to zero outside a large radius and avoid spurious reflections from the boundary by using external complex scaling \cite{Scrinzi2010,Kamban2019,Kamban2020}. See \cref{sec:Wannier} for further details.

\begin{figure*}[t]
	\centering
	\includegraphics[width=1.8\columnwidth]{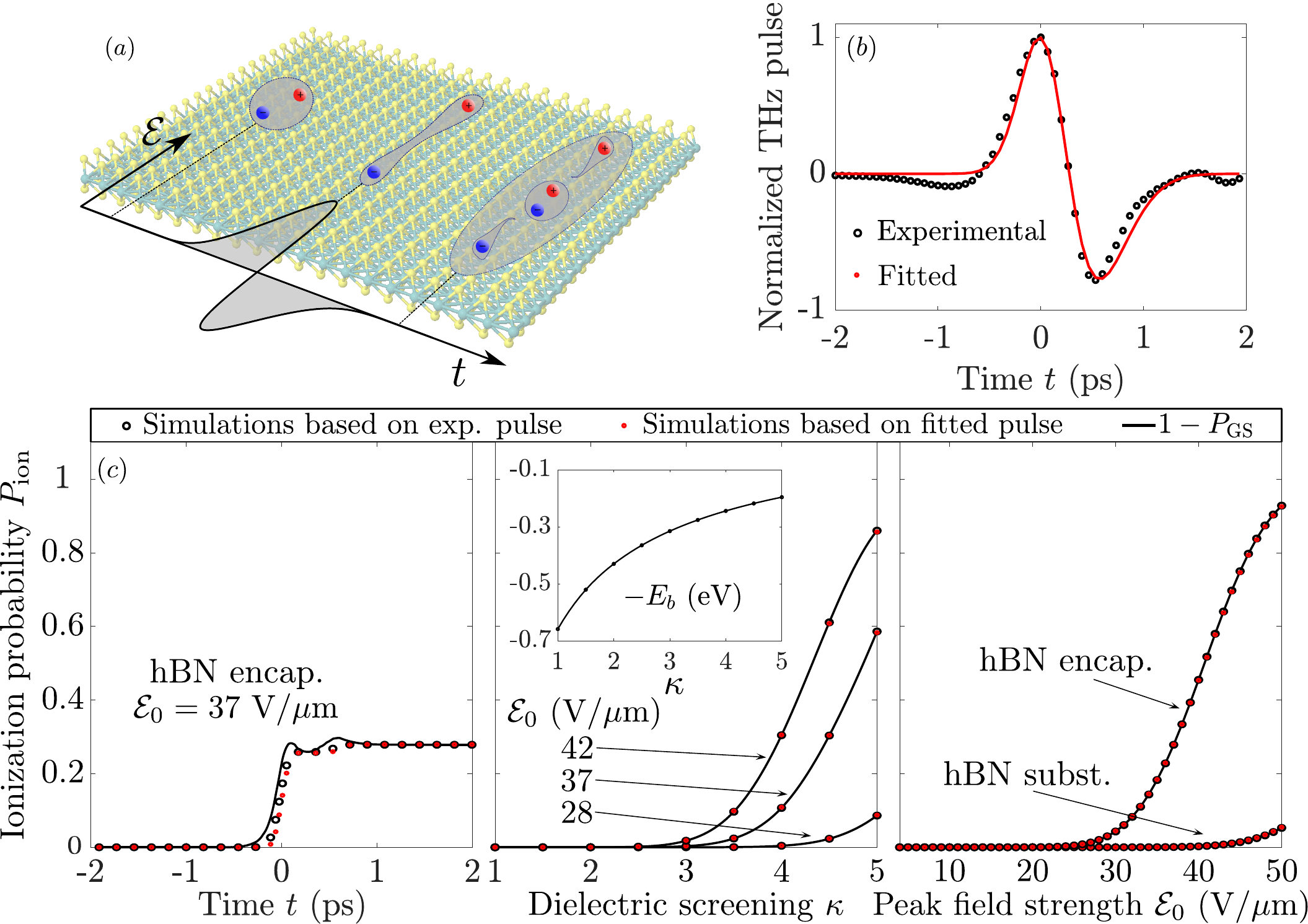}
	\caption{\textbf{(a)} Schematic illustration of an exciton in a THz pulse. The initially bound exciton is polarized as a result of the pulse and is in a superposition of bound and ionized states as the pulse subsides. \textbf{(b)} Traces of the experimental pulse in Ref. \cite{Shi2020} and the fitted pulse in \cref{eq:pulse} are shown as the black circles and red curve, respectively. Figure \textbf{(c)} summarizes the results of an MoS$_2$ exciton subjected to the THz pulses in (b) with the black and red circles representing experimental and fitted pulses, respectively. The solid lines show the probability of depleting the ground state. (left) Time-resolved exciton ionization probability during the THz pulse with a peak field strength of $37$ $\mathrm{V}/\mathrm{\mu m}$ in hBN-encapsulated MoS$_2$. (middle) Exciton ionization probabilities after the pulse for three different field strengths as functions of the dielectric surroundings. The inset shows the corresponding binding energies. (right) Ionization probabilities after the pulse as functions of peak field strength for MoS$_2$ encapsulated by hBN as well as for MoS$_2$ on an hBN substrate. }\label{fig:fig1}
\end{figure*}

\Cref{fig:fig1} (a) shows a schematic illustration of the exciton dynamics during the THz pulse. It may be divided into three temporal regions: (i) the exciton is initially in its unperturbed ground state. (ii) When the pulse is close to its peak field strength, the electron and hole are pulled in opposite directions. It is in this region that the vast majority of ionization occurs. \cite{Langer2018} (iii) As the pulse subsides, the exciton has probabilities $P_{\mathrm{ion}}$ and $1-P_{\mathrm{ion}}$ of being in an ionized and bound state, respectively. Traces of the experimental and fitted THz pulses used in the simulations are shown in \cref{fig:fig1} (b). Here, the black circles represent the experimental THz pulse received from the authors of Ref. \cite{Shi2020}. To ensure easy reproducibility of the results, we introduce a fitted pulse with a known functional form. This pulse is represented by the red line in \cref{fig:fig1} (b), and it is justified by noting that the results obtained using both pulses are in excellent agreement. The functional form is motivated by Ref. \cite{Yang2014} and is given by $\mathcal{E}\left(t\right) = -\frac{\partial A\left(t\right)}{\partial t}\thinspace,$ where the vector potential reads
\begin{align}
	A\left(t\right) = -\frac{\mathcal{E}_0\tau}{c}\exp\left\{\left[\frac{1}{a}\tanh\left(\frac{bt}{\tau}\right)-5^2\right]\frac{t^2}{\tau^2}\right\}\thinspace.\label{eq:pulse}
\end{align}
Here, $\mathcal{E}_0$ is the peak field strength, and $a$ and $b$ dictate the shape of the pulse. In the present paper, we choose $a=0.12$ and $b=3$ to accurately reproduce the main features of the experimental pulse. The coefficient $c$ is chosen such that the peak field strength becomes $\mathcal{E}_0$. We define the pulse duration to be from the moment it reaches $1\%$ of its peak field strength to the moment it again reaches $1\%$ as it subsides. This duration is approximately $\tau$. To approximate the experimental pulse, we initially set $\tau = 2.1$ ps and later investigate changes in $P_{\mathrm{ion}}$ as $\tau$ is varied. Also, in all time-resolved results below, we shift the $t$-axis such that the pulse reaches it peak value at $t=0$.

\section{Results}

\Cref{fig:fig1} (c) summarizes the results for an exciton in MoS$_2$ exposed to a THz pulse. The three panels show the time-resolved ionization probability during the pulse (left), as well as the ionization probability after the pulse as a function of dielectric screening (middle) and peak field strength (right). In these panels, the black and red circles represent ionization probabilities $P_{\mathrm{ion}}$ obtained by using the experimental and fitted THz pulses, respectively, while the solid black lines show the ground state depletion probability $1-P_{\mathrm{GS}}$. The left panel reveals that the dynamics in the fitted and experimental THz pulses agree. Furthermore, while the ground state depletion probability and ionization probability do not coincide during the pulse, they do when the pulse subsides. This indicates that excited states that are completely ionized when the pulse has died out are transiently occupied during the pulse. As will be shown later, this is not the case for shorter pulses, since they are too short to completely ionize the excited states. The dynamics are similar to those in Ref. \cite{Ewers2012}, where exciton occupation probabilities in quantum-well structures subjected to THz radiation were investigated. The middle panel shows ionization probabilities after the pulse has died out for three different peak field strengths as functions of surrounding dielectric screening $\kappa$. We again observe that $1-P_{\mathrm{GS}}$ and $P_{\mathrm{ion}}$ coincide after the pulse. This is a clear indication that the process is adiabatic, and the same conclusion was reached in Ref. \cite{Shi2020}. Comparing the ionization probability with the binding energies for the relevant dielectric screening shown in the inset, we see that the increasing ionization probability is well explained by the reducing binding energy \cite{Kamban2020a,Henriques2020a}. Common dielectric surroundings include SiO$_2$ substrates ($\kappa \approx 2.4)$ and hBN encapsulation ($\kappa \approx 4.5)$.

\begin{figure}[t]
	\centering
	\includegraphics[width=1\columnwidth]{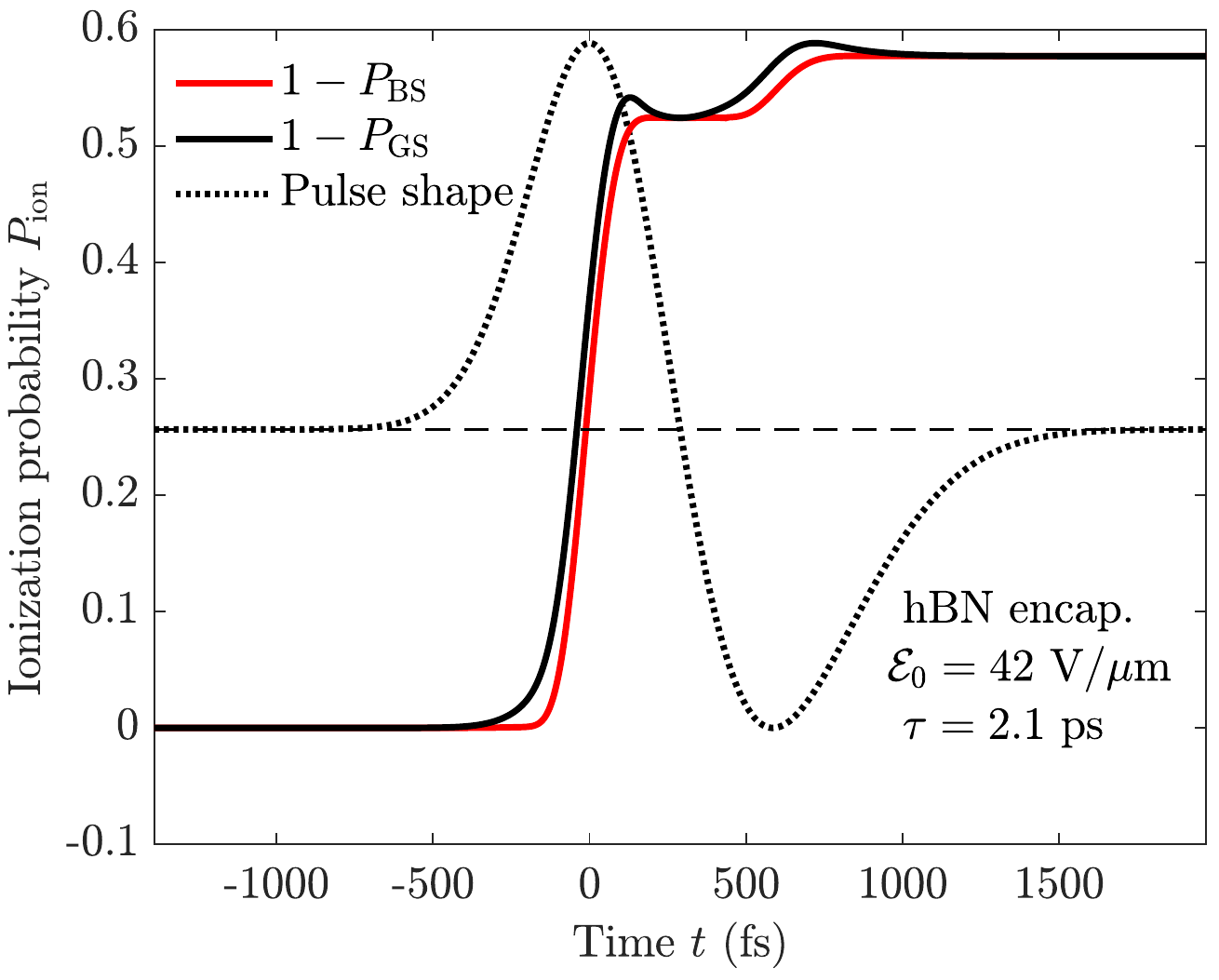}
	\caption{Time-resolved ionization (red) and ground state depletion (black) probabilities for excitons in hBN-encapsulated MoS$_2$. The pulse shape is shown as the dotted line and is centered in time at its peak value of $42$ $\mathrm{V}/\mathrm{\mu m}$ with the dashed line indicating vanishing electric field level. The pulse duration is $\tau = 2.1$ ps.}\label{fig:4}
\end{figure}

The right panel of \cref{fig:fig1} (c) shows ionization probabilities as functions of peak field strength $\mathcal{E}_0$ for MoS$_2$ in two different dielectric surroundings. It is clear that the increased screening from encapsulating the TMD sheet in hBN, as opposed to simply placing it on an hBN substrate, causes a substantial increase in ionization probability. These results suggest that one of the experimental THz pulses used in Ref. \cite{Shi2020} with a peak field strength of $42$ V/$\mu$m would be able to ionize about 58\% of the excitons in hBN-encapsulated MoS$_2$ where almost all of the ionization occurs over an interval of less than $0.5$ ps (see \cref{fig:4}). In their experiment, the authors of Ref. \cite{Shi2020} use a sapphire substrate, which leads to an exciton binding energy of around $240$ meV \cite{Shi2020}. The authors find a broadening of the exciton absorption peak of about $10.8$ meV for a peak field strength of $37$ V/$\mu$m. For our calculations, this binding energy is reproduced by letting $\kappa \approx 4.1$. The middle panel of \cref{fig:fig1} (c) reveals that about $14\%$ of the excitons should be ionized in this case. This corroborates the assumption made by the authors that exciton ionization may be neglected compared to the field-induced phonon broadening. The broadening induced by ionization alone may be estimated by using a static electric field of $37$ V/$\mu$m. In this case, a full-width of about $0.6$ meV is found \cite{Kamban2019}, further confirming that mechanisms other than exciton ionization dominate the broadening of the absorption peaks in this experiment. 

\begin{figure*}[t]
	\centering
	\includegraphics[width=1.8\columnwidth]{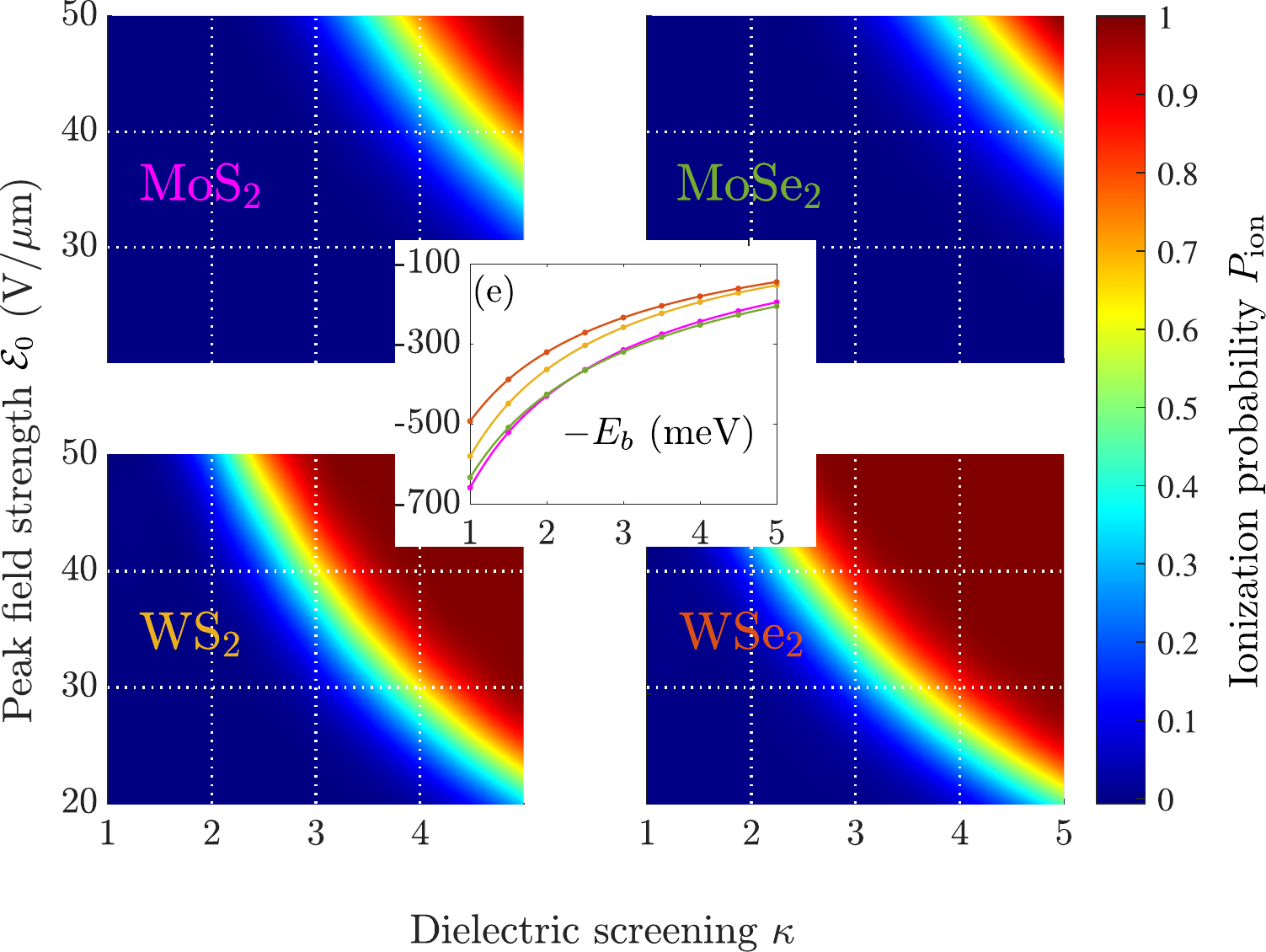}
	\caption{Ionization probability as a function of peak field strength and surrounding dielectric screening for \textbf{(a)} MoS$_2$, \textbf{(b)} MoSe$_2$, \textbf{(c)} WS$_2$, and \textbf{(d)} WSe$_2$. Panel \textbf{(e)} shows the relevant binding energies. The colors of the TMD chemical composition in panels (a-d) correspond to the colors in panel (e). }\label{fig:ionization_prob}
\end{figure*}

The optimal material for a particular device is typically determined by requirements such as sensitivity to a certain frequency range. A prerequisite to material selection is accurate knowledge of their behavior under realistic conditions and in various dielectric environments. We therefore compute the ionization probabilities for four different TMDs using the fitted pulse above, while varying the field strength and dielectric screening. The results are shown as color coded plots in \cref{fig:ionization_prob}. The TMDs considered are (a) MoS$_2$, (b) MoSe$_2$, (c) WS$_2$, and (d) WSe$_2$. The inset shows the exciton binding energies of these TMDs in the relevant range of dielectric surroundings. Common to all TMDs considered, hardly any ionization occurs if either the peak field strength is lower than $20$ V/$\mu$m, or the dielectric screening constant is close to unity (freely suspended). As the field strength and screening increase, an increase in ionization probability is observed. It is immediately clear that the tungsten materials have larger ionization probabilities than the molybdenum materials in the relevant regions for device components. This is due to the lower exciton binding energies in the tungsten materials, which suggests that it is easier to manipulate the free carrier density in these materials. If we consider the case with $\kappa=3$ and $\mathcal{E}_0 = 40$ V/$\mu$m as an example, we see that the TMDs ordered from highest to lowest ionization probability are WSe$_2$, WS$_2$,  MoS$_2$, and  MoSe$_2$. This is in good agreement with the binding energies at $\kappa=3$, for which we have $E_b^{\left(\mathrm{MoSe}_2\right)}>E_b^{\left(\mathrm{MoS}_2\right)}>E_b^{\left(\mathrm{WS}_2\right)}>E_b^{\left(\mathrm{WSe}_2\right)}$.

\begin{figure*}[t]
	\centering
	\includegraphics[width=1.8\columnwidth]{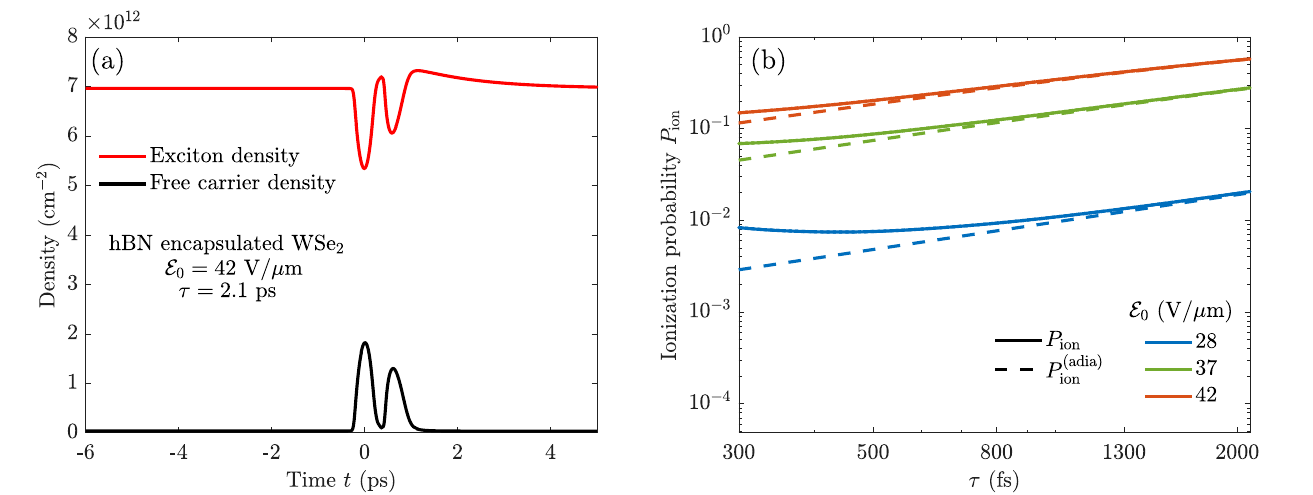}
	\caption{\textbf{(a)} Exciton and free carrier densities in hBN-encapsulated WSe$_2$ obtained by solving the coupled rate equations (see \cref{sec:Free_carrier}). The black and red curves indicate the exciton $n_x$ and free carrier $n$ densities, respectively. The densities are initially in equilibrium, that is $n_x\approx 7\times 10^{12}$ cm$^{-2}$ and $n\approx 3.5\times 10^{10}$ cm$^{-2}$. A THz pulse with a duration $\tau = 2.1$ ps reaching its peak field strength of 42 V/$\mu$m at $t=0$ is later turned on. Realistic results are obtained by using a finite exciton generation rate and taking into account loss mechanisms from impurities (see \cref{sec:Free_carrier}). \textbf{(b)} Ionization probability as a function of pulse duration for various peak field strengths, indicated by the different colors. The solid lines are the ionization $P_{\mathrm{ion}}$ probabilities computed by propagating the Wannier equation in time and projecting on the relevant bound states. The dashed lines are calculated using an adiabatic approach, utilizing the results from static electric fields. The adiabatic probabilities are given by $P^{\left(\mathrm{adia}\right)}_{\mathrm{ion}} = 1 - e^{\tau\Gamma_1}$, where $\Gamma_1 = \{9.62, 154.24, 407.63\}$ ns$^{-1}$ for $\mathcal{E}_0 = \{28,37,42\}$ V/$\mu$m, respectively. See \cref{sec:Adia} for details.}\label{fig:3}
\end{figure*}

To estimate the effect of a THz pulse on the free carrier concentration in a realistic TMD, one must take into account different loss mechanisms such as exciton recombination, exciton-exciton annihilation, and free carrier lifetimes, as well as impact ionization. To accurately explain such effects, sophisticated theoretical modeling is often used \cite{Langer2018}. In the present paper, the focus on obtaining an accurate prediction of ionization probabilities. However, it is interesting to see the effect that exciton ionization has on the free carrier concentration. To this end, we develop a simple model consisting of two coupled rate equations that describe the exciton and free carrier concentrations, respectively. The details may be found in \cref{sec:Free_carrier}. The representative case for hBN-encapsulated WSe$_2$ is shown in \cref{fig:3} (a). Here, a constant exciton generation rate of $G = 2.5\times 10^{24}$ s$^{-1}$cm$^{-2}$ is assumed. The black and red curves represent the exciton and free carrier densities, respectively. The densities are initially assumed to be in equilibrium, and a THz pulse reaching its peak field strength of 42 V/$\mu$m at $t=0$ is then turned on. A rapid increase in free carrier concentration from about $3.5\times 10^{10}$ cm$^{-2}$ at equilibrium to about $1.8\times 10^{12}$ cm$^{-2}$ can be observed. This is an increase in free carrier concentration by a factor of about 51 over a very short interval. After this sudden increase, the concentrations return to their equilibrium values. This ability to dynamically control the charge carrier concentration is very promising for applications such as THz modulators based on monolayer TMDs \cite{Cao2016,Fan2020}. It should also make it possible to measure exciton ionization probabilities by correlating them to free carrier absorption spectra

\section{Shorter pulses}

So far we have only considered THz pulses with duration given by $\tau=2.1$ ps, closely resembling the experimental pulse used in Ref. \cite{Shi2020}. For such large $\tau$ the results are very accurately reproduced by adiabatic calculations, and it is therefore interesting to see how the results differ for shorter pulses. This is summarized in \cref{fig:3} (b), where we show the ionization probability of an exciton in MoS$_2$ encapsulated by hBN after the pulse has died out as a function of the pulse duration. The pulses considered are of the same shape as before but with a shorter duration (and therefore a higher center frequency). Different peak field strengths are indicated by line color. The solid lines show the ionization probabilities computed by propagating the Wannier equation in time (see \cref{sec:Wannier}), while the dashed lines show results from an adiabatic calculation (see \cref{sec:Adia}). We observe that for the longest pulses considered, the adiabatic calculation accurately reproduces the results. Furthermore, we find that the ground state depletion probability and the ionization probability are graphically indistinguishable for all pulse lengths shown in \cref{fig:3} (b). This suggests that the excited states that are occupied during the THz pulse are completely ionized once the pulse dies out. As the duration of the pulse decreases, we observe a deviation from the adiabatic results. This is to be expected, as the coupling between the states and the pulse should cease to be adiabatic for pulses that vary more rapidly with time. The results indicate that if the goal is to ionize as many excitons as possible, a pulse in the THz region is preferable over shorter pulses. The experimental THz pulse in Ref. \cite{Shi2020} corresponds to the longest duration in \cref{fig:3} (b), and therefore seems to be a good candidate for exciton ionization. 

\section*{Summary}

In summary, we have demonstrated using theoretical tools that realistic THz pulses may be used to efficiently ionize excitons in TMD monolayers, a desirable feature in many optoelectronic devices, such as photodetectors and THz modulators. We have shown that for the technologically important case of hBN-encapsulated MoS$_2$, a THz pulse with a peak field strength of 42 V/$\mu$m (as applied in recent experiments \cite{Shi2020}) will ionize about $58\%$ of the photogenerated excitons. Reducing the peak field strength to 30 V/$\mu$m already lowers the ionization probability to about $4.5\%$, revealing the extreme sensitivity to field strength. Similar behavior was demonstrated for MoSe$_2$, whereas exciton ionization probabilities in WS$_2$ and WSe$_2$ were shown to be considerably larger owing to their reduced exciton binding energies. In fact, a peak field of 30 V/$\mu$m should ionize about $96\%$ of the excitons in hBN-encapsulated WSe$_2$. 

By solving the coupled rate equations for the exciton and free carrier concentrations in a realistic 2D semiconductor containing defects, we have demonstrated that THz pulses boost the free carrier concentration considerably over a short time interval, providing dynamical control over both bound and the free charge carriers. For the representative case of hBN-encapsulated WSe$_2$, we found an increase by more than a factor 50. As the free carrier absorption is directly related to the number of free charge carriers, exciton ionization probabilities should be experimentally measurable in these materials by comparing free carrier absorption before and after the THz pulse. Finally, we have shown that reducing the duration of the pulse leads to a process that is no longer adiabatic. 

 \appendix
 \section{WANNIER EXCITONS IN THZ ELECTRIC FIELDS}\label{sec:Wannier}
 
 Within the two-band, effective mass approximation, excitons in a TMD may be described by the two-dimensional Wannier equation \cite{Wannier1937,Lederman1976a}. In terms of the relative coordinate $\boldsymbol{r}$, it reads (atomic units are used throughout)
 \begin{align}\label{eq:Wannier}
 	\left[-\frac{1}{2\mu}\nabla^2 +V\left(\boldsymbol{r}\right)\right]\varphi_{mn}\left(\boldsymbol{r}\right)=E_{mn}\varphi_{mn}\left(\boldsymbol{r}\right)\thinspace,
 \end{align} 
 where $\mu$ is the reduced exciton mass, $V$ the interaction potential, and $m$ and $n$ denote the angular momentum and principle quantum number, respectively. Due to cylindrical symmetry in the unperturbed case, angular mometum is a good quantum number in the Wannier model. In the present paper, we are interested in TMDs surrounded by in-plane-isotropic media with dielectric tensor $\varepsilon_{a/b} = \mathrm{diag}(\varepsilon_x^{\left(a/b\right)},\varepsilon_x^{\left(a/b\right)},\varepsilon_z^{\left(a/b\right)})$ above ($a$) and below ($b$) the sheet, respectively. The dielectric function may be approximated by a linearized form $\varepsilon\left(\boldsymbol{q}\right) = \kappa + r_0q$, where $\boldsymbol{q}$ is the momentum space coordinate, $\kappa = (\sqrt{\varepsilon_x^{\left(a\right)}\varepsilon_z^{\left(a\right)}}+\sqrt{\varepsilon_x^{\left(b\right)}\varepsilon_z^{\left(b\right)}})/2$ is the average dielectric constant between the sub- and superstrate \cite{Trolle2017,Kamban2020b}, and the screening length $r_0$ is related to the polarizability of the TMD monolayer by $r_0 = 2\pi\alpha_{\mathrm{2D}}$. \cite{Cudazzo2011a,Trolle2017} In this case, $V$ is given by the Rytova-Keldysh form \cite{Rytova1967,Keldysh1979,Trolle2017,Kamban2020b}
 \begin{align}\label{eq:Keldysh_pot}
 	V\left(\boldsymbol{r}\right) = -\frac{\pi}{2r_0}\left[\mathrm{H}_0\left(\frac{\kappa r}{r_0}\right)-Y_0\left(\frac{\kappa r}{r_0}\right)\right]\thinspace,
 \end{align}
 where $\mathrm{H}_0$ is the zeroth order Struve function and $Y_0$ the zeroth order Bessel function of the second kind \cite{Abramowitz1972}. We use the experimentally verified values for $\mu$ and $r_0$ from Ref. \cite{Goryca2019}. Additionally, when hBN surroundings are considered, we use the values for $\kappa$ found in Ref. \cite{Goryca2019}.
 
 To study exciton ionization induced by THz pulses, we include a time-dependent electric field term in the Wannier equation. Working in the length gauge under the dipole approximation and letting the field direction be along the $x$-axis, we write
 \begin{align}\label{eq:PerturbedWannier}
 	i\frac{\partial\psi\left(\boldsymbol{r},t\right)}{\partial t}=H(t)\psi\left(\boldsymbol{r},t\right)\thinspace,
 \end{align} 
 with
 \begin{align}\label{eq:TDHamilton}
 	H\left(t\right) = -\frac{1}{2\mu}\nabla^2+ V\left(\boldsymbol{r}\right)+\mathcal{E}\left(t\right)x\thinspace,
 \end{align}
 where $\mathcal{E}\left(t\right)$ is the time-dependent field strength of the THz pulse. We will assume that only the excitonic ground state is occupied initially, and then propagate \cref{eq:PerturbedWannier} in time while computing the ionization probability $P_{\mathrm{ion}}$ as
 \begin{align}
 	P_{\mathrm{ion}}\left(t\right) = 1-\sum_{n,m  \text{ bound}}\left|\braket{\varphi_{mn}\left(\boldsymbol{r}\right)|\psi\left(\boldsymbol{r},t\right)}\right|^2\thinspace,
 \end{align}
 where we only count bound state solutions $\varphi_{mn}$ of \cref{eq:Wannier} (i.e. states with $E_{mn}<0$), and the braket notation has been used to indicate integration with respect to all spatial coordinates. 
 
 To perform numerical computations, the spatial region is truncated by placing the system inside a large radial box such that the wave function is forced to zero outside a radius $R$. This corresponds to introducing an infinite potential outside $R$. As long as $R$ is chosen large enough, the wave function should be well described within this box. Importantly, this allows us to write the wave function as a superposition of the \textit{discrete} eigenstate solutions $\varphi_{mn}$ to the unperturbed problem in \cref{eq:Wannier} subject to $\varphi_{mn}\left(\boldsymbol{r}\right)=0$ for $r\geq R$. We write
 \begin{align}\label{eq:expansion}
 	\psi\left(\boldsymbol{r},t\right) = \sum_{m=0}^M\sum_{n=0}^N c_{mn}\left(t\right)\varphi_{mn}\left(\boldsymbol{r}\right)e^{-iE_{mn}t}\thinspace,
 \end{align}
 with time-dependent expansion coefficients. In practical calculations, we make sure that the results are converged in $M$, $N$, and $R$. Note that, as the electric field is taken along the $x$-direction, only the $\cos\left(m\theta\right)$ angular functions with $m\geq 0$ appear. Substituting \cref{eq:expansion} into \cref{eq:PerturbedWannier}, the following expression may be derived
 \begin{multline}
 	\frac{d}{dt}c_{mn}\left(t\right) = -i\sum_{m'n'}c_{m'n'}\left(t\right)\mathcal{E}\left(t\right)
 	\bra{\varphi_{mn}}x\ket{\varphi_{m'n'}} \\ \times e^{-i\left(E_{m'n'}-E_{mn}\right)t}\thinspace.
 \end{multline}
 Thus, the expansion coefficients may easily be propagated in time, and the occupation probability of state $\varphi_{mn}$ obtained as $|c_{mn}|^2$. The results in the main text have been obtained using an adaptive step-size eight-order Dormand-Prince Runge-Kutta method with embedded fifth- and third-order methods for error control \cite{William2007}. The unperturbed states and energies have been obtained by expanding the eigenstates in a finite element (FE) basis
 \begin{align}
 	\varphi_{mn}\left(\boldsymbol{r}\right) = \sum_{k=1}^{K}\sum_{i=1}^{p}d_i^{\left(m,n,k\right)}\left(t\right)f_i^{\left(k\right)}\left(r\right)\cos\left(m\theta\right)\thinspace,
 \end{align}
 where $k$ and $i$ denote the radial segment and basis function, respectively. One of the advantages with using an FE expansion is that it is easy to implement exterior complex scaling (ECS) outside a particular radius $R_0$. In this region $r\geq R_0$, one lets the radial coordinate rotate into the complex plane. This reduces the amount of spurious reflections from the boundary considerably, and greatly reduces the number of pseudo-continuous states necessary for convergences \cite{Bengtsson2008,Scrinzi2010}. For the exact form of the $f$ functions and further details on the ECS FE calculation, we refer the interested reader to one of our previous papers \cite{Kamban2019}. 
 
 \section{FREE CARRIER DENSITY}\label{sec:Free_carrier}
 
 In this section, we turn to estimating the charge carrier density in hBN-encapsulated WSe$_2$. Without any external perturbations, we may write the initial rate equations as
 \begin{align}\label{eq:nxfree}
 	\frac{dn_x}{dt} =  -\frac{n_x}{\tau_{\mathrm{therm}}} + \gamma_cn^2 - \gamma_Inn_x\thinspace, 
 \end{align}
 \begin{align}\label{eq:nfree}
 	\frac{dn}{dt} =  \frac{n_x}{\tau_{\mathrm{therm}}} - \gamma_c n^2 + \gamma_Inn_x\thinspace,
 \end{align}
 where $n_x$ is the exciton density, $n$ is the density of free carriers, $\tau_{\mathrm{therm}}$ is the exciton ionization rate due to thermal agitation, $\gamma_c$ is the rate at which free electrons and holes combine to form excitons, and $\gamma_I$ is the impact ionization rate, i.e. the rate at which excitons are ionized due to collisions with free carriers. For hBN-encapsulated WSe$_2$, a value of $\tau_{\mathrm{therm}} \approx 100$ ps \cite{Perea-Causin2021} may be used. The impact ionization rate $\gamma_I$ may be estimated by
 \begin{align}
 	\gamma_I = \int_{0}^{\infty}D\left(E\right)v\left(E\right)\sigma_I\left(E\right)dE\thinspace,
 \end{align}
 where $D$ is the energy distribution, $v$ the free carrier velocity, and $\sigma_I$ the exciton impact ionization cross section. In an effective mass approximation, we have $E = k^2/2m_e$ with the effective electron mass $m_e = 0.45$ \cite{Haastrup2018}. We assume the energy distribution to be Maxwellian $D\left(E\right) = \exp\left(-E/k_BT\right)/k_BT$, where $k_B$ is the Boltzmann constant and $T$ the electron temperature. Note that the electron temperature is not necessarily the same as the surrounding temperature. In Ref \cite{Dargys1998} it was found that
 \begin{align}
 	\sigma_I\approx \frac{16}{k_0}\left(\frac{k_0^2}{k^2}-\frac{k_0^4}{k^4}\right)\theta(k-k_0)\thinspace,
 \end{align}
 which has a maximum of $4/k_0$, where $k_0$ is defined in terms of $E_b = k_0^2/2m_e$. Approximating the scattering cross section by its maximum, it is found that
 \begin{align}
 	\gamma_I \approx \frac{1}{m_e}\left[2\sqrt{\frac{\pi}{z}}\Phi_c\left(\sqrt{z}\right)+4e^{-z}\right]\thinspace,
 \end{align}
 where $z=E_b/k_BT$ and $\Phi_c$ is the complementary error function. Using an exciton binding energy of $E_b = 161.4$ meV and assuming the electron temperature is $T=300$ K, a value of $\gamma_I = 0.0043$ cm$^2$/s is found. It should be noted that using a slightly higher temperature does not have a large impact on the value of $\gamma_I$. To estimate $\gamma_c$, we will use the Saha equation which describes the free carrier and exciton densities after statistical equlibration of their chemical potentials in the Boltzmann limit \cite{Kaindl2009}
 \begin{align}\label{eq:Saha}
 	\frac{n^2}{n_x} = \frac{k_BT\mu}{2\pi}e^{-E_b/k_BT}\thinspace.
 \end{align}
 The equilibrium solutions are obtained by setting the first order derivatives in \cref{eq:nxfree,eq:nfree} equal to zero and supplementing with \cref{eq:Saha} to determine $\gamma_c$. Here, we find $\gamma_c \approx 59$ cm$^2$/s. In a realistic system, a laser pump is usually used to generate excitons. There are, furthermore, impurities that lead to finite exciton and free carrier lifetimes. To describe such a system, we write
 \begin{multline}\label{eq:nx}
 	\frac{dn_x}{dt} = G(t) +\left(\frac{d\log P_B}{dt}-\frac{1}{\tau_x} -\frac{1}{\tau_{\mathrm{therm}}}\right)n_x \\ - \gamma_{ee} n_x^2 + \gamma_cn^2 - \gamma_Inn_x\thinspace, 
 \end{multline}
 \begin{align}\label{eq:n}
 	\frac{dn}{dt} = -\frac{n}{\tau_{fc}} + \left(\frac{1}{\tau_{\mathrm{therm}}} - \frac{d\log P_B}{dt}\right)n_x - \gamma_c n^2 + \gamma_Inn_x\thinspace.
 \end{align}
 Here, $G(t)$ is the exciton generation rate, $\tau_x$ is the field-free exciton lifetime, $\gamma_{ee}$ is the exciton-exciton annihilation rate, and $\tau_{fc}$ is the field-free free carrier lifetime. The logarithmic derivative of the total bound-state probability $P_B = \sum_b P_b$ may be understood as a time-dependent ionization rate due to the applied THz pulse. To describe hBN-encapsulated WSe$_2$, we may use $\tau_x \approx 90$ ps \cite{Cadiz2018}, $\gamma_{ee} \approx 0.05$ cm$^2$/s \cite{Massicotte2018, Mouri2014}, and $\tau_{fc} \approx 40$ ps \cite{Aivazian2017} together with the previously found quantities. Here, we also assume a constant exciton generation rate $G = 2.5\times 10^{24}$ s$^{-1}$cm$^{-2}$ due to an external laser pump. The value is chosen such that we obtain the same equilibrium exciton density used in Ref. \cite{Massicotte2018}. We find $n_x\approx 7\times 10^{12}$ cm$^{-2}$ and $n\approx 3.5\times 10^{10}$ cm$^{-2}$. \Cref{eq:nx,eq:n} are then solved starting from these equilibrium solutions. Note that $d\log P_B/dt$ depends on the field strength $\mathcal{E}\left(t\right)$ and is zero when $\mathcal{E}=0$.
 
 \section{ADIABATIC IONIZATION PROBABILITY}\label{sec:Adia}
 
 Assuming that the system is initially in its (non-degenerate) ground state $\varphi_{\mathrm{GS}}$, and that $\mathcal{E}\left(t\right)$ varies sufficiently slowly, the adiabatic theorem states that the system will remain in the eigenstate of $H\left(t\right)$ that evolves from the initial state $\varphi_{\mathrm{GS}}$. This allows us to find the energy at all times by treating time as a parameter, and tracing the relevant state. That is, we find the eigenstates of $H(t')$ for a fixed $t=t'$ at a time
 \begin{align}
 	H\left(t'\right)\varphi_\lambda\left(\boldsymbol{r};t'\right)=E_\lambda\left(t'\right)\varphi_\lambda\left(\boldsymbol{r};t'\right)\thinspace,
 \end{align}
 where $E$ and $\varphi$ depend on $t'$ as a parameter. Then we let $t'$ take on values from the initial to the final $t$, all the while keeping track of the state that evolves from the ground state. Using the complex scaling procedure \cite{Kamban2019}, we are then able to obtain both the real (Stark shift) and imaginary (ionization rate) part of the energy. The relevant (adiabatic) state is
 \begin{align}\label{eq:ad_psi}
 	\psi\left(\boldsymbol{r},t\right) = \exp\left\{-i\int_{-\infty}^tE\left(t'\right)dt'\right\}\varphi_{\mathrm{GS}}\left(\boldsymbol{r};t\right)\thinspace.
 \end{align}
 As the pulse subsides, $\psi$ will return to the ground state $c\varphi_{\mathrm{GS}}\left(\boldsymbol{r}\right)\exp\left(-iE_{\mathrm{GS}}t\right)$, where $c$ is a complex number that arises from the integral in \cref{eq:ad_psi}, resulting in a norm less than unity. Defining the ionization rate as $\Gamma = -2\mathrm{Im}\, E$, we obtain the adiabatic ionization probability
 \begin{align}
 	P_{\mathrm{ion}}^{\left(\mathrm{adia}\right)}\left(t\right) = 1-\exp\left\{-\int_{-\infty}^{t}\Gamma\left(t'\right)dt'\right\}\thinspace.
 \end{align}
 Note that the pulse in Eq. 1 in the main text depends only on $\tau$ as the fraction $t/\tau$. We therefore have
 \begin{align}
 	\Gamma\left(t;\tau,\mathcal{E}_0\right) = \Gamma\left(\frac{t}{\tau};1,\mathcal{E}_0\right)
 \end{align}
 and
 \begin{align}
 	\int_{-\infty}^{\infty}\Gamma\left(t'\right)dt' = \tau \Gamma_1\thinspace,
 \end{align}
 where
 \begin{align}
 	\Gamma_1 = \int_{-\infty}^{\infty}\Gamma\left(t';1,\mathcal{E}_0\right)dt'\thinspace.
 \end{align}
 The adiabatic ionization probability after the pulse is therefore
 \begin{align}
 	P_{\mathrm{ion}}^{\left(\mathrm{adia}\right)}\left(t\to\infty\right) = 1-\exp\left(-\tau \Gamma_1\right)\thinspace.
 \end{align}

\bibliography{../../../../../Litterature/library}

\begin{thebibliography}{63}%
\makeatletter
\providecommand \@ifxundefined [1]{%
 \@ifx{#1\undefined}
}%
\providecommand \@ifnum [1]{%
 \ifnum #1\expandafter \@firstoftwo
 \else \expandafter \@secondoftwo
 \fi
}%
\providecommand \@ifx [1]{%
 \ifx #1\expandafter \@firstoftwo
 \else \expandafter \@secondoftwo
 \fi
}%
\providecommand \natexlab [1]{#1}%
\providecommand \enquote  [1]{``#1''}%
\providecommand \bibnamefont  [1]{#1}%
\providecommand \bibfnamefont [1]{#1}%
\providecommand \citenamefont [1]{#1}%
\providecommand \href@noop [0]{\@secondoftwo}%
\providecommand \href [0]{\begingroup \@sanitize@url \@href}%
\providecommand \@href[1]{\@@startlink{#1}\@@href}%
\providecommand \@@href[1]{\endgroup#1\@@endlink}%
\providecommand \@sanitize@url [0]{\catcode `\\12\catcode `\$12\catcode
  `\&12\catcode `\#12\catcode `\^12\catcode `\_12\catcode `\%12\relax}%
\providecommand \@@startlink[1]{}%
\providecommand \@@endlink[0]{}%
\providecommand \url  [0]{\begingroup\@sanitize@url \@url }%
\providecommand \@url [1]{\endgroup\@href {#1}{\urlprefix }}%
\providecommand \urlprefix  [0]{URL }%
\providecommand \Eprint [0]{\href }%
\providecommand \doibase [0]{https://doi.org/}%
\providecommand \selectlanguage [0]{\@gobble}%
\providecommand \bibinfo  [0]{\@secondoftwo}%
\providecommand \bibfield  [0]{\@secondoftwo}%
\providecommand \translation [1]{[#1]}%
\providecommand \BibitemOpen [0]{}%
\providecommand \bibitemStop [0]{}%
\providecommand \bibitemNoStop [0]{.\EOS\space}%
\providecommand \EOS [0]{\spacefactor3000\relax}%
\providecommand \BibitemShut  [1]{\csname bibitem#1\endcsname}%
\let\auto@bib@innerbib\@empty
\bibitem [{\citenamefont {Schaibley}\ \emph {et~al.}(2016)\citenamefont
  {Schaibley}, \citenamefont {Yu}, \citenamefont {Clark}, \citenamefont
  {Rivera}, \citenamefont {Ross}, \citenamefont {Seyler}, \citenamefont {Yao},\
  and\ \citenamefont {Xu}}]{Schaibley2016}%
  \BibitemOpen
  \bibfield  {author} {\bibinfo {author} {\bibfnamefont {J.~R.}\ \bibnamefont
  {Schaibley}}, \bibinfo {author} {\bibfnamefont {H.}~\bibnamefont {Yu}},
  \bibinfo {author} {\bibfnamefont {G.}~\bibnamefont {Clark}}, \bibinfo
  {author} {\bibfnamefont {P.}~\bibnamefont {Rivera}}, \bibinfo {author}
  {\bibfnamefont {J.~S.}\ \bibnamefont {Ross}}, \bibinfo {author}
  {\bibfnamefont {K.~L.}\ \bibnamefont {Seyler}}, \bibinfo {author}
  {\bibfnamefont {W.}~\bibnamefont {Yao}},\ and\ \bibinfo {author}
  {\bibfnamefont {X.}~\bibnamefont {Xu}},\ }\bibfield  {title} {\bibinfo
  {title} {{Valleytronics in 2D materials}},\ }\href
  {https://doi.org/10.1038/natrevmats.2016.55} {\bibfield  {journal} {\bibinfo
  {journal} {Nat. Rev. Mater.}\ }\textbf {\bibinfo {volume} {1}},\ \bibinfo
  {pages} {16055} (\bibinfo {year} {2016})}\BibitemShut {NoStop}%
\bibitem [{\citenamefont {Gong}\ \emph {et~al.}(2017)\citenamefont {Gong},
  \citenamefont {Zhang}, \citenamefont {Chen}, \citenamefont {Chu},
  \citenamefont {Lei}, \citenamefont {Pu}, \citenamefont {Dai}, \citenamefont
  {Wu}, \citenamefont {Cheng}, \citenamefont {Zhai}, \citenamefont {Li},\ and\
  \citenamefont {Xiong}}]{Gong2017}%
  \BibitemOpen
  \bibfield  {author} {\bibinfo {author} {\bibfnamefont {C.}~\bibnamefont
  {Gong}}, \bibinfo {author} {\bibfnamefont {Y.}~\bibnamefont {Zhang}},
  \bibinfo {author} {\bibfnamefont {W.}~\bibnamefont {Chen}}, \bibinfo {author}
  {\bibfnamefont {J.}~\bibnamefont {Chu}}, \bibinfo {author} {\bibfnamefont
  {T.}~\bibnamefont {Lei}}, \bibinfo {author} {\bibfnamefont {J.}~\bibnamefont
  {Pu}}, \bibinfo {author} {\bibfnamefont {L.}~\bibnamefont {Dai}}, \bibinfo
  {author} {\bibfnamefont {C.}~\bibnamefont {Wu}}, \bibinfo {author}
  {\bibfnamefont {Y.}~\bibnamefont {Cheng}}, \bibinfo {author} {\bibfnamefont
  {T.}~\bibnamefont {Zhai}}, \bibinfo {author} {\bibfnamefont {L.}~\bibnamefont
  {Li}},\ and\ \bibinfo {author} {\bibfnamefont {J.}~\bibnamefont {Xiong}},\
  }\bibfield  {title} {\bibinfo {title} {{Electronic and optoelectronic
  applications based on 2D novel anisotropic transition metal
  dichalcogenides}},\ }\href {https://doi.org/10.1002/advs.201700231}
  {\bibfield  {journal} {\bibinfo  {journal} {Adv. Sci.}\ }\textbf {\bibinfo
  {volume} {4}},\ \bibinfo {pages} {1700231} (\bibinfo {year}
  {2017})}\BibitemShut {NoStop}%
\bibitem [{\citenamefont {Yin}\ \emph {et~al.}(2012)\citenamefont {Yin},
  \citenamefont {Li}, \citenamefont {Li}, \citenamefont {Jiang}, \citenamefont
  {Shi}, \citenamefont {Sun}, \citenamefont {Lu}, \citenamefont {Zhang},
  \citenamefont {Chen},\ and\ \citenamefont {Zhang}}]{Yin2012a}%
  \BibitemOpen
  \bibfield  {author} {\bibinfo {author} {\bibfnamefont {Z.}~\bibnamefont
  {Yin}}, \bibinfo {author} {\bibfnamefont {H.}~\bibnamefont {Li}}, \bibinfo
  {author} {\bibfnamefont {H.}~\bibnamefont {Li}}, \bibinfo {author}
  {\bibfnamefont {L.}~\bibnamefont {Jiang}}, \bibinfo {author} {\bibfnamefont
  {Y.}~\bibnamefont {Shi}}, \bibinfo {author} {\bibfnamefont {Y.}~\bibnamefont
  {Sun}}, \bibinfo {author} {\bibfnamefont {G.}~\bibnamefont {Lu}}, \bibinfo
  {author} {\bibfnamefont {Q.}~\bibnamefont {Zhang}}, \bibinfo {author}
  {\bibfnamefont {X.}~\bibnamefont {Chen}},\ and\ \bibinfo {author}
  {\bibfnamefont {H.}~\bibnamefont {Zhang}},\ }\bibfield  {title} {\bibinfo
  {title} {{Single-layer MoS2 phototransistors}},\ }\href
  {https://doi.org/10.1021/nn2024557} {\bibfield  {journal} {\bibinfo
  {journal} {ACS Nano}\ }\textbf {\bibinfo {volume} {6}},\ \bibinfo {pages}
  {74} (\bibinfo {year} {2012})}\BibitemShut {NoStop}%
\bibitem [{\citenamefont {Lopez-Sanchez}\ \emph {et~al.}(2013)\citenamefont
  {Lopez-Sanchez}, \citenamefont {Lembke}, \citenamefont {Kayci}, \citenamefont
  {Radenovic},\ and\ \citenamefont {Kis}}]{Lopez2013}%
  \BibitemOpen
  \bibfield  {author} {\bibinfo {author} {\bibfnamefont {O.}~\bibnamefont
  {Lopez-Sanchez}}, \bibinfo {author} {\bibfnamefont {D.}~\bibnamefont
  {Lembke}}, \bibinfo {author} {\bibfnamefont {M.}~\bibnamefont {Kayci}},
  \bibinfo {author} {\bibfnamefont {A.}~\bibnamefont {Radenovic}},\ and\
  \bibinfo {author} {\bibfnamefont {A.}~\bibnamefont {Kis}},\ }\bibfield
  {title} {\bibinfo {title} {{Ultrasensitive photodetectors based on monolayer
  MoS2}},\ }\href {https://doi.org/10.1038/nnano.2013.100} {\bibfield
  {journal} {\bibinfo  {journal} {Nat. Nanotechnol.}\ }\textbf {\bibinfo
  {volume} {8}},\ \bibinfo {pages} {497} (\bibinfo {year} {2013})}\BibitemShut
  {NoStop}%
\bibitem [{\citenamefont {Wang}\ \emph {et~al.}(2015)\citenamefont {Wang},
  \citenamefont {Zhang}, \citenamefont {Chan}, \citenamefont {Tiwari},\ and\
  \citenamefont {Rana}}]{Wang2015a}%
  \BibitemOpen
  \bibfield  {author} {\bibinfo {author} {\bibfnamefont {H.}~\bibnamefont
  {Wang}}, \bibinfo {author} {\bibfnamefont {C.}~\bibnamefont {Zhang}},
  \bibinfo {author} {\bibfnamefont {W.}~\bibnamefont {Chan}}, \bibinfo {author}
  {\bibfnamefont {S.}~\bibnamefont {Tiwari}},\ and\ \bibinfo {author}
  {\bibfnamefont {F.}~\bibnamefont {Rana}},\ }\bibfield  {title} {\bibinfo
  {title} {{Ultrafast response of monolayer molybdenum disulfide
  photodetectors}},\ }\href {https://doi.org/10.1038/ncomms9831} {\bibfield
  {journal} {\bibinfo  {journal} {Nat. Commun.}\ }\textbf {\bibinfo {volume}
  {6}},\ \bibinfo {pages} {8831} (\bibinfo {year} {2015})}\BibitemShut
  {NoStop}%
\bibitem [{\citenamefont {Cao}\ \emph {et~al.}(2016)\citenamefont {Cao},
  \citenamefont {Gan}, \citenamefont {Geng}, \citenamefont {Liu}, \citenamefont
  {Yang}, \citenamefont {Bao},\ and\ \citenamefont {Chen}}]{Cao2016}%
  \BibitemOpen
  \bibfield  {author} {\bibinfo {author} {\bibfnamefont {Y.}~\bibnamefont
  {Cao}}, \bibinfo {author} {\bibfnamefont {S.}~\bibnamefont {Gan}}, \bibinfo
  {author} {\bibfnamefont {Z.}~\bibnamefont {Geng}}, \bibinfo {author}
  {\bibfnamefont {J.}~\bibnamefont {Liu}}, \bibinfo {author} {\bibfnamefont
  {Y.}~\bibnamefont {Yang}}, \bibinfo {author} {\bibfnamefont {Q.}~\bibnamefont
  {Bao}},\ and\ \bibinfo {author} {\bibfnamefont {H.}~\bibnamefont {Chen}},\
  }\bibfield  {title} {\bibinfo {title} {{Optically tuned terahertz modulator
  based on annealed multilayer MoS2}},\ }\href
  {https://doi.org/10.1038/srep22899} {\bibfield  {journal} {\bibinfo
  {journal} {Sci. Rep.}\ }\textbf {\bibinfo {volume} {6}},\ \bibinfo {pages}
  {22899} (\bibinfo {year} {2016})}\BibitemShut {NoStop}%
\bibitem [{\citenamefont {Fan}\ \emph {et~al.}(2020)\citenamefont {Fan},
  \citenamefont {Geng}, \citenamefont {Fang}, \citenamefont {Lv}, \citenamefont
  {Su}, \citenamefont {Wang}, \citenamefont {Liu},\ and\ \citenamefont
  {Chen}}]{Fan2020}%
  \BibitemOpen
  \bibfield  {author} {\bibinfo {author} {\bibfnamefont {Z.}~\bibnamefont
  {Fan}}, \bibinfo {author} {\bibfnamefont {Z.}~\bibnamefont {Geng}}, \bibinfo
  {author} {\bibfnamefont {W.}~\bibnamefont {Fang}}, \bibinfo {author}
  {\bibfnamefont {X.}~\bibnamefont {Lv}}, \bibinfo {author} {\bibfnamefont
  {Y.}~\bibnamefont {Su}}, \bibinfo {author} {\bibfnamefont {S.}~\bibnamefont
  {Wang}}, \bibinfo {author} {\bibfnamefont {J.}~\bibnamefont {Liu}},\ and\
  \bibinfo {author} {\bibfnamefont {H.}~\bibnamefont {Chen}},\ }\bibfield
  {title} {\bibinfo {title} {{Characteristics of transition metal
  dichalcogenides in optical pumped modulator of terahertz wave}},\ }\href
  {https://doi.org/10.1063/1.5141511} {\bibfield  {journal} {\bibinfo
  {journal} {AIP Adv.}\ }\textbf {\bibinfo {volume} {10}},\ \bibinfo {pages}
  {045304} (\bibinfo {year} {2020})}\BibitemShut {NoStop}%
\bibitem [{\citenamefont {Bernardi}\ \emph {et~al.}(2013)\citenamefont
  {Bernardi}, \citenamefont {Palummo},\ and\ \citenamefont
  {Grossman}}]{Bernardi2013}%
  \BibitemOpen
  \bibfield  {author} {\bibinfo {author} {\bibfnamefont {M.}~\bibnamefont
  {Bernardi}}, \bibinfo {author} {\bibfnamefont {M.}~\bibnamefont {Palummo}},\
  and\ \bibinfo {author} {\bibfnamefont {J.~C.}\ \bibnamefont {Grossman}},\
  }\bibfield  {title} {\bibinfo {title} {{Extraordinary sunlight absorption and
  one nanometer thick photovoltaics using two-dimensional monolayer
  materials}},\ }\href {https://doi.org/10.1021/nl401544y} {\bibfield
  {journal} {\bibinfo  {journal} {Nano Lett.}\ }\textbf {\bibinfo {volume}
  {13}},\ \bibinfo {pages} {3664} (\bibinfo {year} {2013})}\BibitemShut
  {NoStop}%
\bibitem [{\citenamefont {Lopez-Sanchez}\ \emph {et~al.}(2014)\citenamefont
  {Lopez-Sanchez}, \citenamefont {{Alarcon Llado}}, \citenamefont {Koman},
  \citenamefont {{Fontcuberta I Morral}}, \citenamefont {Radenovic},\ and\
  \citenamefont {Kis}}]{Lopez-Sanchez2014}%
  \BibitemOpen
  \bibfield  {author} {\bibinfo {author} {\bibfnamefont {O.}~\bibnamefont
  {Lopez-Sanchez}}, \bibinfo {author} {\bibfnamefont {E.}~\bibnamefont
  {{Alarcon Llado}}}, \bibinfo {author} {\bibfnamefont {V.}~\bibnamefont
  {Koman}}, \bibinfo {author} {\bibfnamefont {A.}~\bibnamefont {{Fontcuberta I
  Morral}}}, \bibinfo {author} {\bibfnamefont {A.}~\bibnamefont {Radenovic}},\
  and\ \bibinfo {author} {\bibfnamefont {A.}~\bibnamefont {Kis}},\ }\bibfield
  {title} {\bibinfo {title} {{Light generation and harvesting in a van der
  waals heterostructure}},\ }\href {https://doi.org/10.1021/nn500480u}
  {\bibfield  {journal} {\bibinfo  {journal} {ACS Nano}\ }\textbf {\bibinfo
  {volume} {8}},\ \bibinfo {pages} {3042} (\bibinfo {year} {2014})}\BibitemShut
  {NoStop}%
\bibitem [{\citenamefont {Ramasubramaniam}(2012)}]{Ramasubramaniam2012}%
  \BibitemOpen
  \bibfield  {author} {\bibinfo {author} {\bibfnamefont {A.}~\bibnamefont
  {Ramasubramaniam}},\ }\bibfield  {title} {\bibinfo {title} {{Large excitonic
  effects in monolayers of molybdenum and tungsten dichalcogenides}},\ }\href
  {https://doi.org/10.1103/PhysRevB.86.115409} {\bibfield  {journal} {\bibinfo
  {journal} {Phys. Rev. B}\ }\textbf {\bibinfo {volume} {86}},\ \bibinfo
  {pages} {115409} (\bibinfo {year} {2012})}\BibitemShut {NoStop}%
\bibitem [{\citenamefont {Berkelbach}\ \emph {et~al.}(2013)\citenamefont
  {Berkelbach}, \citenamefont {Hybertsen},\ and\ \citenamefont
  {Reichman}}]{Berkelbach2013}%
  \BibitemOpen
  \bibfield  {author} {\bibinfo {author} {\bibfnamefont {T.~C.}\ \bibnamefont
  {Berkelbach}}, \bibinfo {author} {\bibfnamefont {M.~S.}\ \bibnamefont
  {Hybertsen}},\ and\ \bibinfo {author} {\bibfnamefont {D.~R.}\ \bibnamefont
  {Reichman}},\ }\bibfield  {title} {\bibinfo {title} {{Theory of neutral and
  charged excitons in monolayer transition metal dichalcogenides}},\ }\href
  {https://doi.org/10.1103/PhysRevB.88.045318} {\bibfield  {journal} {\bibinfo
  {journal} {Phys. Rev. B}\ }\textbf {\bibinfo {volume} {88}},\ \bibinfo
  {pages} {45318} (\bibinfo {year} {2013})}\BibitemShut {NoStop}%
\bibitem [{\citenamefont {Olsen}\ \emph {et~al.}(2016)\citenamefont {Olsen},
  \citenamefont {Latini}, \citenamefont {Rasmussen},\ and\ \citenamefont
  {Thygesen}}]{Olsen2016}%
  \BibitemOpen
  \bibfield  {author} {\bibinfo {author} {\bibfnamefont {T.}~\bibnamefont
  {Olsen}}, \bibinfo {author} {\bibfnamefont {S.}~\bibnamefont {Latini}},
  \bibinfo {author} {\bibfnamefont {F.}~\bibnamefont {Rasmussen}},\ and\
  \bibinfo {author} {\bibfnamefont {K.~S.}\ \bibnamefont {Thygesen}},\
  }\bibfield  {title} {\bibinfo {title} {{Simple Screened Hydrogen Model of
  Excitons in Two-Dimensional Materials}},\ }\href
  {https://doi.org/10.1103/PhysRevLett.116.056401} {\bibfield  {journal}
  {\bibinfo  {journal} {Phys. Rev. Lett.}\ }\textbf {\bibinfo {volume} {116}},\
  \bibinfo {pages} {56401} (\bibinfo {year} {2016})}\BibitemShut {NoStop}%
\bibitem [{\citenamefont {Wang}\ \emph {et~al.}(2012)\citenamefont {Wang},
  \citenamefont {Kalantar-Zadeh}, \citenamefont {Kis}, \citenamefont
  {Coleman},\ and\ \citenamefont {Strano}}]{Wang2012b}%
  \BibitemOpen
  \bibfield  {author} {\bibinfo {author} {\bibfnamefont {Q.~H.}\ \bibnamefont
  {Wang}}, \bibinfo {author} {\bibfnamefont {K.}~\bibnamefont
  {Kalantar-Zadeh}}, \bibinfo {author} {\bibfnamefont {A.}~\bibnamefont {Kis}},
  \bibinfo {author} {\bibfnamefont {J.~N.}\ \bibnamefont {Coleman}},\ and\
  \bibinfo {author} {\bibfnamefont {M.~S.}\ \bibnamefont {Strano}},\ }\bibfield
   {title} {\bibinfo {title} {{Electronics and optoelectronics of
  two-dimensional transition metal dichalcogenides}},\ }\href
  {https://doi.org/10.1038/nnano.2012.193} {\bibfield  {journal} {\bibinfo
  {journal} {Nat. Nanotechnol.}\ }\textbf {\bibinfo {volume} {7}},\ \bibinfo
  {pages} {699} (\bibinfo {year} {2012})}\BibitemShut {NoStop}%
\bibitem [{\citenamefont {Qiu}\ \emph {et~al.}(2013)\citenamefont {Qiu},
  \citenamefont {da~Jornada},\ and\ \citenamefont {Louie}}]{Qiu2013}%
  \BibitemOpen
  \bibfield  {author} {\bibinfo {author} {\bibfnamefont {D.~Y.}\ \bibnamefont
  {Qiu}}, \bibinfo {author} {\bibfnamefont {F.~H.}\ \bibnamefont
  {da~Jornada}},\ and\ \bibinfo {author} {\bibfnamefont {S.~G.}\ \bibnamefont
  {Louie}},\ }\bibfield  {title} {\bibinfo {title} {{Optical spectrum of MoS2:
  Many-body effects and diversity of exciton states}},\ }\href
  {https://doi.org/10.1103/PhysRevLett.111.216805} {\bibfield  {journal}
  {\bibinfo  {journal} {Phys. Rev. Lett.}\ }\textbf {\bibinfo {volume} {111}},\
  \bibinfo {pages} {216805} (\bibinfo {year} {2013})}\BibitemShut {NoStop}%
\bibitem [{\citenamefont {Trolle}\ \emph {et~al.}(2014)\citenamefont {Trolle},
  \citenamefont {Seifert},\ and\ \citenamefont {Pedersen}}]{Trolle2014}%
  \BibitemOpen
  \bibfield  {author} {\bibinfo {author} {\bibfnamefont {M.~L.}\ \bibnamefont
  {Trolle}}, \bibinfo {author} {\bibfnamefont {G.}~\bibnamefont {Seifert}},\
  and\ \bibinfo {author} {\bibfnamefont {T.~G.}\ \bibnamefont {Pedersen}},\
  }\bibfield  {title} {\bibinfo {title} {{Theory of excitonic second-harmonic
  generation in monolayer MoS2}},\ }\href
  {https://doi.org/10.1103/PhysRevB.89.235410} {\bibfield  {journal} {\bibinfo
  {journal} {Phys. Rev. B}\ }\textbf {\bibinfo {volume} {89}},\ \bibinfo
  {pages} {235410} (\bibinfo {year} {2014})}\BibitemShut {NoStop}%
\bibitem [{\citenamefont {Haastrup}\ \emph {et~al.}(2016)\citenamefont
  {Haastrup}, \citenamefont {Latini}, \citenamefont {Bolotin},\ and\
  \citenamefont {Thygesen}}]{Haastrup2016}%
  \BibitemOpen
  \bibfield  {author} {\bibinfo {author} {\bibfnamefont {S.}~\bibnamefont
  {Haastrup}}, \bibinfo {author} {\bibfnamefont {S.}~\bibnamefont {Latini}},
  \bibinfo {author} {\bibfnamefont {K.}~\bibnamefont {Bolotin}},\ and\ \bibinfo
  {author} {\bibfnamefont {K.~S.}\ \bibnamefont {Thygesen}},\ }\bibfield
  {title} {\bibinfo {title} {{Stark shift and electric-field-induced
  dissociation of excitons in monolayer MoS2 and hBN/MoS2 heterostructures}},\
  }\href {https://doi.org/10.1103/PhysRevB.94.041401} {\bibfield  {journal}
  {\bibinfo  {journal} {Phys. Rev. B}\ }\textbf {\bibinfo {volume} {94}},\
  \bibinfo {pages} {041401} (\bibinfo {year} {2016})}\BibitemShut {NoStop}%
\bibitem [{\citenamefont {Scharf}\ \emph {et~al.}(2016)\citenamefont {Scharf},
  \citenamefont {Frank}, \citenamefont {Gmitra}, \citenamefont {Fabian},
  \citenamefont {{\v{Z}}uti{\'{c}}},\ and\ \citenamefont
  {Perebeinos}}]{Scharf2016a}%
  \BibitemOpen
  \bibfield  {author} {\bibinfo {author} {\bibfnamefont {B.}~\bibnamefont
  {Scharf}}, \bibinfo {author} {\bibfnamefont {T.}~\bibnamefont {Frank}},
  \bibinfo {author} {\bibfnamefont {M.}~\bibnamefont {Gmitra}}, \bibinfo
  {author} {\bibfnamefont {J.}~\bibnamefont {Fabian}}, \bibinfo {author}
  {\bibfnamefont {I.}~\bibnamefont {{\v{Z}}uti{\'{c}}}},\ and\ \bibinfo
  {author} {\bibfnamefont {V.}~\bibnamefont {Perebeinos}},\ }\bibfield  {title}
  {\bibinfo {title} {{Excitonic Stark effect in MoS2 monolayers}},\ }\href
  {https://doi.org/10.1103/PhysRevB.94.245434} {\bibfield  {journal} {\bibinfo
  {journal} {Phys. Rev. B}\ }\textbf {\bibinfo {volume} {94}},\ \bibinfo
  {pages} {245434} (\bibinfo {year} {2016})}\BibitemShut {NoStop}%
\bibitem [{\citenamefont {Massicotte}\ \emph {et~al.}(2018)\citenamefont
  {Massicotte}, \citenamefont {Vialla}, \citenamefont {Schmidt}, \citenamefont
  {Lundeberg}, \citenamefont {Latini}, \citenamefont {Haastrup}, \citenamefont
  {Danovich}, \citenamefont {Davydovskaya}, \citenamefont {Watanabe},
  \citenamefont {Taniguchi}, \citenamefont {Fal'ko}, \citenamefont {Thygesen},
  \citenamefont {Pedersen},\ and\ \citenamefont {Koppens}}]{Massicotte2018}%
  \BibitemOpen
  \bibfield  {author} {\bibinfo {author} {\bibfnamefont {M.}~\bibnamefont
  {Massicotte}}, \bibinfo {author} {\bibfnamefont {F.}~\bibnamefont {Vialla}},
  \bibinfo {author} {\bibfnamefont {P.}~\bibnamefont {Schmidt}}, \bibinfo
  {author} {\bibfnamefont {M.~B.}\ \bibnamefont {Lundeberg}}, \bibinfo {author}
  {\bibfnamefont {S.}~\bibnamefont {Latini}}, \bibinfo {author} {\bibfnamefont
  {S.}~\bibnamefont {Haastrup}}, \bibinfo {author} {\bibfnamefont
  {M.}~\bibnamefont {Danovich}}, \bibinfo {author} {\bibfnamefont
  {D.}~\bibnamefont {Davydovskaya}}, \bibinfo {author} {\bibfnamefont
  {K.}~\bibnamefont {Watanabe}}, \bibinfo {author} {\bibfnamefont
  {T.}~\bibnamefont {Taniguchi}}, \bibinfo {author} {\bibfnamefont {V.~I.}\
  \bibnamefont {Fal'ko}}, \bibinfo {author} {\bibfnamefont {K.~S.}\
  \bibnamefont {Thygesen}}, \bibinfo {author} {\bibfnamefont {T.~G.}\
  \bibnamefont {Pedersen}},\ and\ \bibinfo {author} {\bibfnamefont {F.~H.}\
  \bibnamefont {Koppens}},\ }\bibfield  {title} {\bibinfo {title}
  {{Dissociation of two-dimensional excitons in monolayer WSe2}},\ }\href
  {https://doi.org/10.1038/s41467-018-03864-y} {\bibfield  {journal} {\bibinfo
  {journal} {Nat. Commun.}\ }\textbf {\bibinfo {volume} {9}},\ \bibinfo {pages}
  {1633} (\bibinfo {year} {2018})}\BibitemShut {NoStop}%
\bibitem [{\citenamefont {Klots}\ \emph {et~al.}(2014)\citenamefont {Klots},
  \citenamefont {Newaz}, \citenamefont {Wang}, \citenamefont {Prasai},
  \citenamefont {Krzyzanowska}, \citenamefont {Lin}, \citenamefont {Caudel},
  \citenamefont {Ghimire}, \citenamefont {Yan}, \citenamefont {Ivanov},
  \citenamefont {Velizhanin}, \citenamefont {Burger}, \citenamefont {Mandrus},
  \citenamefont {Tolk}, \citenamefont {Pantelides},\ and\ \citenamefont
  {Bolotin}}]{Klots2014}%
  \BibitemOpen
  \bibfield  {author} {\bibinfo {author} {\bibfnamefont {A.~R.}\ \bibnamefont
  {Klots}}, \bibinfo {author} {\bibfnamefont {A.~K.}\ \bibnamefont {Newaz}},
  \bibinfo {author} {\bibfnamefont {B.}~\bibnamefont {Wang}}, \bibinfo {author}
  {\bibfnamefont {D.}~\bibnamefont {Prasai}}, \bibinfo {author} {\bibfnamefont
  {H.}~\bibnamefont {Krzyzanowska}}, \bibinfo {author} {\bibfnamefont
  {J.}~\bibnamefont {Lin}}, \bibinfo {author} {\bibfnamefont {D.}~\bibnamefont
  {Caudel}}, \bibinfo {author} {\bibfnamefont {N.~J.}\ \bibnamefont {Ghimire}},
  \bibinfo {author} {\bibfnamefont {J.}~\bibnamefont {Yan}}, \bibinfo {author}
  {\bibfnamefont {B.~L.}\ \bibnamefont {Ivanov}}, \bibinfo {author}
  {\bibfnamefont {K.~A.}\ \bibnamefont {Velizhanin}}, \bibinfo {author}
  {\bibfnamefont {A.}~\bibnamefont {Burger}}, \bibinfo {author} {\bibfnamefont
  {D.~G.}\ \bibnamefont {Mandrus}}, \bibinfo {author} {\bibfnamefont {N.~H.}\
  \bibnamefont {Tolk}}, \bibinfo {author} {\bibfnamefont {S.~T.}\ \bibnamefont
  {Pantelides}},\ and\ \bibinfo {author} {\bibfnamefont {K.~I.}\ \bibnamefont
  {Bolotin}},\ }\bibfield  {title} {\bibinfo {title} {{Probing excitonic states
  in suspended two-dimensional semiconductors by photocurrent spectroscopy}},\
  }\href {https://doi.org/10.1038/srep06608} {\bibfield  {journal} {\bibinfo
  {journal} {Sci. Rep.}\ }\textbf {\bibinfo {volume} {4}},\ \bibinfo {pages}
  {6608} (\bibinfo {year} {2014})}\BibitemShut {NoStop}%
\bibitem [{\citenamefont {Kamban}\ and\ \citenamefont
  {Pedersen}(2019)}]{Kamban2019}%
  \BibitemOpen
  \bibfield  {author} {\bibinfo {author} {\bibfnamefont {H.~C.}\ \bibnamefont
  {Kamban}}\ and\ \bibinfo {author} {\bibfnamefont {T.~G.}\ \bibnamefont
  {Pedersen}},\ }\bibfield  {title} {\bibinfo {title} {{Field-induced
  dissociation of two-dimensional excitons in transition metal
  dichalcogenides}},\ }\href {https://doi.org/10.1103/PhysRevB.100.045307}
  {\bibfield  {journal} {\bibinfo  {journal} {Phys. Rev. B}\ }\textbf {\bibinfo
  {volume} {100}},\ \bibinfo {pages} {045307} (\bibinfo {year}
  {2019})}\BibitemShut {NoStop}%
\bibitem [{\citenamefont {Pedersen}\ \emph {et~al.}(2016)\citenamefont
  {Pedersen}, \citenamefont {Latini}, \citenamefont {Thygesen}, \citenamefont
  {Mera},\ and\ \citenamefont {Nikoli{\'{c}}}}]{Pedersen2016b}%
  \BibitemOpen
  \bibfield  {author} {\bibinfo {author} {\bibfnamefont {T.~G.}\ \bibnamefont
  {Pedersen}}, \bibinfo {author} {\bibfnamefont {S.}~\bibnamefont {Latini}},
  \bibinfo {author} {\bibfnamefont {K.~S.}\ \bibnamefont {Thygesen}}, \bibinfo
  {author} {\bibfnamefont {H.}~\bibnamefont {Mera}},\ and\ \bibinfo {author}
  {\bibfnamefont {B.~K.}\ \bibnamefont {Nikoli{\'{c}}}},\ }\bibfield  {title}
  {\bibinfo {title} {{Exciton ionization in multilayer transition-metal
  dichalcogenides}},\ }\href {https://doi.org/10.1088/1367-2630/18/7/073043}
  {\bibfield  {journal} {\bibinfo  {journal} {New J. Phys.}\ }\textbf {\bibinfo
  {volume} {18}},\ \bibinfo {pages} {073043} (\bibinfo {year}
  {2016})}\BibitemShut {NoStop}%
\bibitem [{\citenamefont {Kamban}\ and\ \citenamefont
  {Pedersen}(2020)}]{Kamban2020a}%
  \BibitemOpen
  \bibfield  {author} {\bibinfo {author} {\bibfnamefont {H.~C.}\ \bibnamefont
  {Kamban}}\ and\ \bibinfo {author} {\bibfnamefont {T.~G.}\ \bibnamefont
  {Pedersen}},\ }\bibfield  {title} {\bibinfo {title} {{Interlayer excitons in
  van der Waals heterostructures: Binding energy, Stark shift, and
  field-induced dissociation}},\ }\href
  {https://doi.org/10.1038/s41598-020-62431-y} {\bibfield  {journal} {\bibinfo
  {journal} {Sci. Rep.}\ }\textbf {\bibinfo {volume} {10}},\ \bibinfo {pages}
  {5537} (\bibinfo {year} {2020})}\BibitemShut {NoStop}%
\bibitem [{\citenamefont {Jones}\ \emph {et~al.}(1993)\citenamefont {Jones},
  \citenamefont {You},\ and\ \citenamefont {Bucksbaum}}]{Jones1993}%
  \BibitemOpen
  \bibfield  {author} {\bibinfo {author} {\bibfnamefont {R.~R.}\ \bibnamefont
  {Jones}}, \bibinfo {author} {\bibfnamefont {D.}~\bibnamefont {You}},\ and\
  \bibinfo {author} {\bibfnamefont {P.~H.}\ \bibnamefont {Bucksbaum}},\
  }\bibfield  {title} {\bibinfo {title} {{Ionization of Rydberg atoms by
  subpicosecond half-cycle electromagnetic pulses}},\ }\href
  {https://doi.org/10.1103/PhysRevLett.76.2436} {\bibfield  {journal} {\bibinfo
   {journal} {Phys. Rev. Lett.}\ }\textbf {\bibinfo {volume} {70}},\ \bibinfo
  {pages} {1236} (\bibinfo {year} {1993})}\BibitemShut {NoStop}%
\bibitem [{\citenamefont {Li}\ and\ \citenamefont {Jones}(2014)}]{Li2014b}%
  \BibitemOpen
  \bibfield  {author} {\bibinfo {author} {\bibfnamefont {S.}~\bibnamefont
  {Li}}\ and\ \bibinfo {author} {\bibfnamefont {R.~R.}\ \bibnamefont {Jones}},\
  }\bibfield  {title} {\bibinfo {title} {{Ionization of excited atoms by
  intense single-cycle THz pulses}},\ }\href
  {https://doi.org/10.1103/PhysRevLett.112.143006} {\bibfield  {journal}
  {\bibinfo  {journal} {Phys. Rev. Lett.}\ }\textbf {\bibinfo {volume} {112}},\
  \bibinfo {pages} {143006} (\bibinfo {year} {2014})}\BibitemShut {NoStop}%
\bibitem [{\citenamefont {Ewers}\ \emph {et~al.}(2012)\citenamefont {Ewers},
  \citenamefont {K{\"{o}}ster}, \citenamefont {Woscholski}, \citenamefont
  {Koch}, \citenamefont {Chatterjee}, \citenamefont {Khitrova}, \citenamefont
  {Gibbs}, \citenamefont {Klettke}, \citenamefont {Kira},\ and\ \citenamefont
  {Koch}}]{Ewers2012}%
  \BibitemOpen
  \bibfield  {author} {\bibinfo {author} {\bibfnamefont {B.}~\bibnamefont
  {Ewers}}, \bibinfo {author} {\bibfnamefont {N.~S.}\ \bibnamefont
  {K{\"{o}}ster}}, \bibinfo {author} {\bibfnamefont {R.}~\bibnamefont
  {Woscholski}}, \bibinfo {author} {\bibfnamefont {M.}~\bibnamefont {Koch}},
  \bibinfo {author} {\bibfnamefont {S.}~\bibnamefont {Chatterjee}}, \bibinfo
  {author} {\bibfnamefont {G.}~\bibnamefont {Khitrova}}, \bibinfo {author}
  {\bibfnamefont {H.~M.}\ \bibnamefont {Gibbs}}, \bibinfo {author}
  {\bibfnamefont {A.~C.}\ \bibnamefont {Klettke}}, \bibinfo {author}
  {\bibfnamefont {M.}~\bibnamefont {Kira}},\ and\ \bibinfo {author}
  {\bibfnamefont {S.~W.}\ \bibnamefont {Koch}},\ }\bibfield  {title} {\bibinfo
  {title} {{Ionization of coherent excitons by strong terahertz fields}},\
  }\href {https://doi.org/10.1103/PhysRevB.85.075307} {\bibfield  {journal}
  {\bibinfo  {journal} {Phys. Rev. B}\ }\textbf {\bibinfo {volume} {85}},\
  \bibinfo {pages} {075307} (\bibinfo {year} {2012})}\BibitemShut {NoStop}%
\bibitem [{\citenamefont {Stein}\ \emph {et~al.}(2018)\citenamefont {Stein},
  \citenamefont {Lammers}, \citenamefont {Steiner}, \citenamefont {Richter},
  \citenamefont {Koch}, \citenamefont {Koch},\ and\ \citenamefont
  {Kira}}]{Stein2018a}%
  \BibitemOpen
  \bibfield  {author} {\bibinfo {author} {\bibfnamefont {M.}~\bibnamefont
  {Stein}}, \bibinfo {author} {\bibfnamefont {C.}~\bibnamefont {Lammers}},
  \bibinfo {author} {\bibfnamefont {J.~T.}\ \bibnamefont {Steiner}}, \bibinfo
  {author} {\bibfnamefont {P.~H.}\ \bibnamefont {Richter}}, \bibinfo {author}
  {\bibfnamefont {S.~W.}\ \bibnamefont {Koch}}, \bibinfo {author}
  {\bibfnamefont {M.}~\bibnamefont {Koch}},\ and\ \bibinfo {author}
  {\bibfnamefont {M.}~\bibnamefont {Kira}},\ }\bibfield  {title} {\bibinfo
  {title} {{Exciton ionization by THz pulses in germanium}},\ }\href
  {https://doi.org/10.1088/1361-6455/aabac7} {\bibfield  {journal} {\bibinfo
  {journal} {J. Phys. B At. Mol. Opt. Phys.}\ }\textbf {\bibinfo {volume}
  {51}},\ \bibinfo {pages} {154001} (\bibinfo {year} {2018})}\BibitemShut
  {NoStop}%
\bibitem [{\citenamefont {Murotani}\ \emph {et~al.}(2018)\citenamefont
  {Murotani}, \citenamefont {Takayama}, \citenamefont {Sekiguchi},
  \citenamefont {Kim}, \citenamefont {Akiyama},\ and\ \citenamefont
  {Shimano}}]{Murotani2018}%
  \BibitemOpen
  \bibfield  {author} {\bibinfo {author} {\bibfnamefont {Y.}~\bibnamefont
  {Murotani}}, \bibinfo {author} {\bibfnamefont {M.}~\bibnamefont {Takayama}},
  \bibinfo {author} {\bibfnamefont {F.}~\bibnamefont {Sekiguchi}}, \bibinfo
  {author} {\bibfnamefont {C.}~\bibnamefont {Kim}}, \bibinfo {author}
  {\bibfnamefont {H.}~\bibnamefont {Akiyama}},\ and\ \bibinfo {author}
  {\bibfnamefont {R.}~\bibnamefont {Shimano}},\ }\bibfield  {title} {\bibinfo
  {title} {{Terahertz field-induced ionization and perturbed free induction
  decay of excitons in bulk GaAs}},\ }\href
  {https://doi.org/10.1088/1361-6463/aaa989} {\bibfield  {journal} {\bibinfo
  {journal} {J. Phys. D. Appl. Phys.}\ }\textbf {\bibinfo {volume} {51}},\
  \bibinfo {pages} {114001} (\bibinfo {year} {2018})}\BibitemShut {NoStop}%
\bibitem [{\citenamefont {Klein}\ \emph {et~al.}(2016)\citenamefont {Klein},
  \citenamefont {Wierzbowski}, \citenamefont {Regler}, \citenamefont {Becker},
  \citenamefont {Heimbach}, \citenamefont {M{\"{u}}ller}, \citenamefont
  {Kaniber},\ and\ \citenamefont {Finley}}]{Klein2016}%
  \BibitemOpen
  \bibfield  {author} {\bibinfo {author} {\bibfnamefont {J.}~\bibnamefont
  {Klein}}, \bibinfo {author} {\bibfnamefont {J.}~\bibnamefont {Wierzbowski}},
  \bibinfo {author} {\bibfnamefont {A.}~\bibnamefont {Regler}}, \bibinfo
  {author} {\bibfnamefont {J.}~\bibnamefont {Becker}}, \bibinfo {author}
  {\bibfnamefont {F.}~\bibnamefont {Heimbach}}, \bibinfo {author}
  {\bibfnamefont {K.}~\bibnamefont {M{\"{u}}ller}}, \bibinfo {author}
  {\bibfnamefont {M.}~\bibnamefont {Kaniber}},\ and\ \bibinfo {author}
  {\bibfnamefont {J.~J.}\ \bibnamefont {Finley}},\ }\bibfield  {title}
  {\bibinfo {title} {{Stark effect spectroscopy of mono- and few-layer MoS2}},\
  }\href {https://doi.org/10.1021/acs.nanolett.5b03954} {\bibfield  {journal}
  {\bibinfo  {journal} {Nano Lett.}\ }\textbf {\bibinfo {volume} {16}},\
  \bibinfo {pages} {1554} (\bibinfo {year} {2016})}\BibitemShut {NoStop}%
\bibitem [{\citenamefont {Pedersen}(2016)}]{Pedersen2016a}%
  \BibitemOpen
  \bibfield  {author} {\bibinfo {author} {\bibfnamefont {T.~G.}\ \bibnamefont
  {Pedersen}},\ }\bibfield  {title} {\bibinfo {title} {{Exciton Stark shift and
  electroabsorption in monolayer transition-metal dichalcogenides}},\ }\href
  {https://doi.org/10.1103/PhysRevB.94.125424} {\bibfield  {journal} {\bibinfo
  {journal} {Phys. Rev. B}\ }\textbf {\bibinfo {volume} {94}},\ \bibinfo
  {pages} {125424} (\bibinfo {year} {2016})}\BibitemShut {NoStop}%
\bibitem [{\citenamefont {Cavalcante}\ \emph {et~al.}(2018)\citenamefont
  {Cavalcante}, \citenamefont {{Da Costa}}, \citenamefont {Farias},
  \citenamefont {Reichman},\ and\ \citenamefont {Chaves}}]{Cavalcante2018}%
  \BibitemOpen
  \bibfield  {author} {\bibinfo {author} {\bibfnamefont {L.~S.}\ \bibnamefont
  {Cavalcante}}, \bibinfo {author} {\bibfnamefont {D.~R.}\ \bibnamefont {{Da
  Costa}}}, \bibinfo {author} {\bibfnamefont {G.~A.}\ \bibnamefont {Farias}},
  \bibinfo {author} {\bibfnamefont {D.~R.}\ \bibnamefont {Reichman}},\ and\
  \bibinfo {author} {\bibfnamefont {A.}~\bibnamefont {Chaves}},\ }\bibfield
  {title} {\bibinfo {title} {{Stark shift of excitons and trions in
  two-dimensional materials}},\ }\href
  {https://doi.org/10.1103/PhysRevB.98.245309} {\bibfield  {journal} {\bibinfo
  {journal} {Phys. Rev. B}\ }\textbf {\bibinfo {volume} {98}},\ \bibinfo
  {pages} {245309} (\bibinfo {year} {2018})}\BibitemShut {NoStop}%
\bibitem [{\citenamefont {Dow}\ and\ \citenamefont {Redfield}(1970)}]{Dow1970}%
  \BibitemOpen
  \bibfield  {author} {\bibinfo {author} {\bibfnamefont {J.~D.}\ \bibnamefont
  {Dow}}\ and\ \bibinfo {author} {\bibfnamefont {D.}~\bibnamefont {Redfield}},\
  }\bibfield  {title} {\bibinfo {title} {{Electroabsorption in semiconductors:
  The excitonic absorption edge}},\ }\href
  {https://doi.org/10.1103/PhysRevB.1.3358} {\bibfield  {journal} {\bibinfo
  {journal} {Phys. Rev. B}\ }\textbf {\bibinfo {volume} {1}},\ \bibinfo {pages}
  {3358} (\bibinfo {year} {1970})}\BibitemShut {NoStop}%
\bibitem [{\citenamefont {Miller}\ \emph {et~al.}(1985)\citenamefont {Miller},
  \citenamefont {Chemla}, \citenamefont {Damen}, \citenamefont {Gossard},
  \citenamefont {Wiegmann}, \citenamefont {Wood},\ and\ \citenamefont
  {Burrus}}]{Miller1985}%
  \BibitemOpen
  \bibfield  {author} {\bibinfo {author} {\bibfnamefont {D.~A.~B.}\
  \bibnamefont {Miller}}, \bibinfo {author} {\bibfnamefont {D.~S.}\
  \bibnamefont {Chemla}}, \bibinfo {author} {\bibfnamefont {T.~C.}\
  \bibnamefont {Damen}}, \bibinfo {author} {\bibfnamefont {A.~C.}\ \bibnamefont
  {Gossard}}, \bibinfo {author} {\bibfnamefont {W.}~\bibnamefont {Wiegmann}},
  \bibinfo {author} {\bibfnamefont {T.~H.}\ \bibnamefont {Wood}},\ and\
  \bibinfo {author} {\bibfnamefont {C.~A.}\ \bibnamefont {Burrus}},\ }\bibfield
   {title} {\bibinfo {title} {{Electric field dependence of optical absorption
  near the band gap of quantum-well structures}},\ }\href
  {https://doi.org/10.1103/PhysRevB.32.1043} {\bibfield  {journal} {\bibinfo
  {journal} {Phys. Rev. B}\ }\textbf {\bibinfo {volume} {32}},\ \bibinfo
  {pages} {1043} (\bibinfo {year} {1985})}\BibitemShut {NoStop}%
\bibitem [{\citenamefont {Shi}\ \emph {et~al.}(2020)\citenamefont {Shi},
  \citenamefont {Baldini}, \citenamefont {Latini}, \citenamefont {Sato},
  \citenamefont {Zhang}, \citenamefont {Pein}, \citenamefont {Shen},
  \citenamefont {Kong}, \citenamefont {Rubio}, \citenamefont {Gedik},\ and\
  \citenamefont {Nelson}}]{Shi2020}%
  \BibitemOpen
  \bibfield  {author} {\bibinfo {author} {\bibfnamefont {J.}~\bibnamefont
  {Shi}}, \bibinfo {author} {\bibfnamefont {E.}~\bibnamefont {Baldini}},
  \bibinfo {author} {\bibfnamefont {S.}~\bibnamefont {Latini}}, \bibinfo
  {author} {\bibfnamefont {S.~A.}\ \bibnamefont {Sato}}, \bibinfo {author}
  {\bibfnamefont {Y.}~\bibnamefont {Zhang}}, \bibinfo {author} {\bibfnamefont
  {B.~C.}\ \bibnamefont {Pein}}, \bibinfo {author} {\bibfnamefont {P.~C.}\
  \bibnamefont {Shen}}, \bibinfo {author} {\bibfnamefont {J.}~\bibnamefont
  {Kong}}, \bibinfo {author} {\bibfnamefont {A.}~\bibnamefont {Rubio}},
  \bibinfo {author} {\bibfnamefont {N.}~\bibnamefont {Gedik}},\ and\ \bibinfo
  {author} {\bibfnamefont {K.~A.}\ \bibnamefont {Nelson}},\ }\bibfield  {title}
  {\bibinfo {title} {{Room temperature terahertz electroabsorption modulation
  by excitons in monolayer transition metal dichalcogenides}},\ }\href
  {https://doi.org/10.1021/acs.nanolett.0c01134} {\bibfield  {journal}
  {\bibinfo  {journal} {Nano Lett.}\ }\textbf {\bibinfo {volume} {20}},\
  \bibinfo {pages} {5214} (\bibinfo {year} {2020})}\BibitemShut {NoStop}%
\bibitem [{\citenamefont {Ogawa}\ \emph {et~al.}(2010)\citenamefont {Ogawa},
  \citenamefont {Watanabe}, \citenamefont {Minami},\ and\ \citenamefont
  {Shimano}}]{Ogawa2010}%
  \BibitemOpen
  \bibfield  {author} {\bibinfo {author} {\bibfnamefont {T.}~\bibnamefont
  {Ogawa}}, \bibinfo {author} {\bibfnamefont {S.}~\bibnamefont {Watanabe}},
  \bibinfo {author} {\bibfnamefont {N.}~\bibnamefont {Minami}},\ and\ \bibinfo
  {author} {\bibfnamefont {R.}~\bibnamefont {Shimano}},\ }\bibfield  {title}
  {\bibinfo {title} {{Room temperature terahertz electro-optic modulation by
  excitons in carbon nanotubes}},\ }\href {https://doi.org/10.1063/1.3470105}
  {\bibfield  {journal} {\bibinfo  {journal} {Appl. Phys. Lett.}\ }\textbf
  {\bibinfo {volume} {97}},\ \bibinfo {pages} {041111} (\bibinfo {year}
  {2010})}\BibitemShut {NoStop}%
\bibitem [{\citenamefont {Yuan}\ and\ \citenamefont {Huang}(2015)}]{Yuan2015}%
  \BibitemOpen
  \bibfield  {author} {\bibinfo {author} {\bibfnamefont {L.}~\bibnamefont
  {Yuan}}\ and\ \bibinfo {author} {\bibfnamefont {L.}~\bibnamefont {Huang}},\
  }\bibfield  {title} {\bibinfo {title} {{Exciton dynamics and annihilation in
  WS2 2D semiconductors}},\ }\href {https://doi.org/10.1039/c5nr00383k}
  {\bibfield  {journal} {\bibinfo  {journal} {Nanoscale}\ }\textbf {\bibinfo
  {volume} {7}},\ \bibinfo {pages} {7402} (\bibinfo {year} {2015})}\BibitemShut
  {NoStop}%
\bibitem [{\citenamefont {Cadiz}\ \emph {et~al.}(2018)\citenamefont {Cadiz},
  \citenamefont {Robert}, \citenamefont {Courtade}, \citenamefont {Manca},
  \citenamefont {Martinelli}, \citenamefont {Taniguchi}, \citenamefont
  {Watanabe}, \citenamefont {Amand}, \citenamefont {Rowe}, \citenamefont
  {Paget}, \citenamefont {Urbaszek},\ and\ \citenamefont {Marie}}]{Cadiz2018}%
  \BibitemOpen
  \bibfield  {author} {\bibinfo {author} {\bibfnamefont {F.}~\bibnamefont
  {Cadiz}}, \bibinfo {author} {\bibfnamefont {C.}~\bibnamefont {Robert}},
  \bibinfo {author} {\bibfnamefont {E.}~\bibnamefont {Courtade}}, \bibinfo
  {author} {\bibfnamefont {M.}~\bibnamefont {Manca}}, \bibinfo {author}
  {\bibfnamefont {L.}~\bibnamefont {Martinelli}}, \bibinfo {author}
  {\bibfnamefont {T.}~\bibnamefont {Taniguchi}}, \bibinfo {author}
  {\bibfnamefont {K.}~\bibnamefont {Watanabe}}, \bibinfo {author}
  {\bibfnamefont {T.}~\bibnamefont {Amand}}, \bibinfo {author} {\bibfnamefont
  {A.~C.}\ \bibnamefont {Rowe}}, \bibinfo {author} {\bibfnamefont
  {D.}~\bibnamefont {Paget}}, \bibinfo {author} {\bibfnamefont
  {B.}~\bibnamefont {Urbaszek}},\ and\ \bibinfo {author} {\bibfnamefont
  {X.}~\bibnamefont {Marie}},\ }\bibfield  {title} {\bibinfo {title} {{Exciton
  diffusion in WSe2 monolayers embedded in a van der Waals heterostructure}},\
  }\href {https://doi.org/10.1063/1.5026478} {\bibfield  {journal} {\bibinfo
  {journal} {Appl. Phys. Lett.}\ }\textbf {\bibinfo {volume} {112}},\ \bibinfo
  {pages} {152106} (\bibinfo {year} {2018})}\BibitemShut {NoStop}%
\bibitem [{\citenamefont {Mouri}\ \emph {et~al.}(2014)\citenamefont {Mouri},
  \citenamefont {Miyauchi}, \citenamefont {Toh}, \citenamefont {Zhao},
  \citenamefont {Eda},\ and\ \citenamefont {Matsuda}}]{Mouri2014}%
  \BibitemOpen
  \bibfield  {author} {\bibinfo {author} {\bibfnamefont {S.}~\bibnamefont
  {Mouri}}, \bibinfo {author} {\bibfnamefont {Y.}~\bibnamefont {Miyauchi}},
  \bibinfo {author} {\bibfnamefont {M.}~\bibnamefont {Toh}}, \bibinfo {author}
  {\bibfnamefont {W.}~\bibnamefont {Zhao}}, \bibinfo {author} {\bibfnamefont
  {G.}~\bibnamefont {Eda}},\ and\ \bibinfo {author} {\bibfnamefont
  {K.}~\bibnamefont {Matsuda}},\ }\bibfield  {title} {\bibinfo {title}
  {{Nonlinear photoluminescence in atomically thin layered WSe2 arising from
  diffusion-assisted exciton-exciton annihilation}},\ }\href
  {https://doi.org/10.1103/PhysRevB.90.155449} {\bibfield  {journal} {\bibinfo
  {journal} {Phys. Rev. B}\ }\textbf {\bibinfo {volume} {90}},\ \bibinfo
  {pages} {155449} (\bibinfo {year} {2014})}\BibitemShut {NoStop}%
\bibitem [{\citenamefont {Shi}\ \emph {et~al.}(2013)\citenamefont {Shi},
  \citenamefont {Yan}, \citenamefont {Bertolazzi}, \citenamefont {Brivio},
  \citenamefont {Gao}, \citenamefont {Kis}, \citenamefont {Jena}, \citenamefont
  {Xing},\ and\ \citenamefont {Huang}}]{Shi2013a}%
  \BibitemOpen
  \bibfield  {author} {\bibinfo {author} {\bibfnamefont {H.}~\bibnamefont
  {Shi}}, \bibinfo {author} {\bibfnamefont {R.}~\bibnamefont {Yan}}, \bibinfo
  {author} {\bibfnamefont {S.}~\bibnamefont {Bertolazzi}}, \bibinfo {author}
  {\bibfnamefont {J.}~\bibnamefont {Brivio}}, \bibinfo {author} {\bibfnamefont
  {B.}~\bibnamefont {Gao}}, \bibinfo {author} {\bibfnamefont {A.}~\bibnamefont
  {Kis}}, \bibinfo {author} {\bibfnamefont {D.}~\bibnamefont {Jena}}, \bibinfo
  {author} {\bibfnamefont {H.~G.}\ \bibnamefont {Xing}},\ and\ \bibinfo
  {author} {\bibfnamefont {L.}~\bibnamefont {Huang}},\ }\bibfield  {title}
  {\bibinfo {title} {{Exciton dynamics in suspended monolayer and few-layer
  MoS2 2D crystals}},\ }\href {https://doi.org/10.1021/nn303973r} {\bibfield
  {journal} {\bibinfo  {journal} {ACS Nano}\ }\textbf {\bibinfo {volume} {7}},\
  \bibinfo {pages} {1072} (\bibinfo {year} {2013})}\BibitemShut {NoStop}%
\bibitem [{\citenamefont {Korn}\ \emph {et~al.}(2011)\citenamefont {Korn},
  \citenamefont {Heydrich}, \citenamefont {Hirmer}, \citenamefont
  {Schmutzler},\ and\ \citenamefont {Schller}}]{Korn2011}%
  \BibitemOpen
  \bibfield  {author} {\bibinfo {author} {\bibfnamefont {T.}~\bibnamefont
  {Korn}}, \bibinfo {author} {\bibfnamefont {S.}~\bibnamefont {Heydrich}},
  \bibinfo {author} {\bibfnamefont {M.}~\bibnamefont {Hirmer}}, \bibinfo
  {author} {\bibfnamefont {J.}~\bibnamefont {Schmutzler}},\ and\ \bibinfo
  {author} {\bibfnamefont {C.}~\bibnamefont {Schller}},\ }\bibfield  {title}
  {\bibinfo {title} {{Low-temperature photocarrier dynamics in monolayer
  MoS2}},\ }\href {https://doi.org/10.1063/1.3636402} {\bibfield  {journal}
  {\bibinfo  {journal} {Appl. Phys. Lett.}\ }\textbf {\bibinfo {volume} {99}},\
  \bibinfo {pages} {102109} (\bibinfo {year} {2011})}\BibitemShut {NoStop}%
\bibitem [{\citenamefont {Cudazzo}\ \emph {et~al.}(2011)\citenamefont
  {Cudazzo}, \citenamefont {Tokatly},\ and\ \citenamefont
  {Rubio}}]{Cudazzo2011a}%
  \BibitemOpen
  \bibfield  {author} {\bibinfo {author} {\bibfnamefont {P.}~\bibnamefont
  {Cudazzo}}, \bibinfo {author} {\bibfnamefont {I.~V.}\ \bibnamefont
  {Tokatly}},\ and\ \bibinfo {author} {\bibfnamefont {A.}~\bibnamefont
  {Rubio}},\ }\bibfield  {title} {\bibinfo {title} {{Dielectric screening in
  two-dimensional insulators: Implications for excitonic and impurity states in
  graphane}},\ }\href {https://doi.org/10.1103/PhysRevB.84.085406} {\bibfield
  {journal} {\bibinfo  {journal} {Phys. Rev. B}\ }\textbf {\bibinfo {volume}
  {84}},\ \bibinfo {pages} {085406} (\bibinfo {year} {2011})}\BibitemShut
  {NoStop}%
\bibitem [{\citenamefont {Trolle}\ \emph {et~al.}(2017)\citenamefont {Trolle},
  \citenamefont {Pedersen},\ and\ \citenamefont {V{\'{e}}niard}}]{Trolle2017}%
  \BibitemOpen
  \bibfield  {author} {\bibinfo {author} {\bibfnamefont {M.~L.}\ \bibnamefont
  {Trolle}}, \bibinfo {author} {\bibfnamefont {T.~G.}\ \bibnamefont
  {Pedersen}},\ and\ \bibinfo {author} {\bibfnamefont {V.}~\bibnamefont
  {V{\'{e}}niard}},\ }\bibfield  {title} {\bibinfo {title} {{Model dielectric
  function for 2D semiconductors including substrate screening}},\ }\href
  {https://doi.org/srep39844} {\bibfield  {journal} {\bibinfo  {journal} {Sci.
  Rep.}\ }\textbf {\bibinfo {volume} {7}},\ \bibinfo {pages} {39844} (\bibinfo
  {year} {2017})}\BibitemShut {NoStop}%
\bibitem [{\citenamefont {Rytova}(1967)}]{Rytova1967}%
  \BibitemOpen
  \bibfield  {author} {\bibinfo {author} {\bibfnamefont {N.~S.}\ \bibnamefont
  {Rytova}},\ }\bibfield  {title} {\bibinfo {title} {{The screened potential of
  a point charge in a thin film}},\ }\href@noop {} {\bibfield  {journal}
  {\bibinfo  {journal} {Moscow Univ. Phys. Bull.}\ }\textbf {\bibinfo {volume}
  {3}},\ \bibinfo {pages} {18} (\bibinfo {year} {1967})}\BibitemShut {NoStop}%
\bibitem [{\citenamefont {Keldysh}(1979)}]{Keldysh1979}%
  \BibitemOpen
  \bibfield  {author} {\bibinfo {author} {\bibfnamefont {L.~V.}\ \bibnamefont
  {Keldysh}},\ }\bibfield  {title} {\bibinfo {title} {{Coulomb interaction in
  thin semiconductor and semimetal films}},\ }\href@noop {} {\bibfield
  {journal} {\bibinfo  {journal} {JETP Lett.}\ }\textbf {\bibinfo {volume}
  {29}},\ \bibinfo {pages} {658} (\bibinfo {year} {1979})}\BibitemShut
  {NoStop}%
\bibitem [{\citenamefont {Kamban}\ \emph
  {et~al.}(2020{\natexlab{a}})\citenamefont {Kamban}, \citenamefont
  {Pedersen},\ and\ \citenamefont {Peres}}]{Kamban2020b}%
  \BibitemOpen
  \bibfield  {author} {\bibinfo {author} {\bibfnamefont {H.~C.}\ \bibnamefont
  {Kamban}}, \bibinfo {author} {\bibfnamefont {T.~G.}\ \bibnamefont
  {Pedersen}},\ and\ \bibinfo {author} {\bibfnamefont {N.~M.~R.}\ \bibnamefont
  {Peres}},\ }\bibfield  {title} {\bibinfo {title} {{Anisotropic Stark shift,
  field-induced dissociation, and electroabsorption of excitons in
  phosphorene}},\ }\href {https://doi.org/10.1103/physrevb.102.115305}
  {\bibfield  {journal} {\bibinfo  {journal} {Phys. Rev. B}\ }\textbf {\bibinfo
  {volume} {102}},\ \bibinfo {pages} {115305} (\bibinfo {year}
  {2020}{\natexlab{a}})}\BibitemShut {NoStop}%
\bibitem [{\citenamefont {Wannier}(1937)}]{Wannier1937}%
  \BibitemOpen
  \bibfield  {author} {\bibinfo {author} {\bibfnamefont {G.~H.}\ \bibnamefont
  {Wannier}},\ }\bibfield  {title} {\bibinfo {title} {{The structure of
  electronic excitation levels in insulating crystals}},\ }\href
  {https://doi.org/10.1103/PhysRev.52.191} {\bibfield  {journal} {\bibinfo
  {journal} {Phys. Rev.}\ }\textbf {\bibinfo {volume} {52}},\ \bibinfo {pages}
  {191} (\bibinfo {year} {1937})}\BibitemShut {NoStop}%
\bibitem [{\citenamefont {Lederman}\ and\ \citenamefont
  {Dow}(1976)}]{Lederman1976a}%
  \BibitemOpen
  \bibfield  {author} {\bibinfo {author} {\bibfnamefont {F.~L.}\ \bibnamefont
  {Lederman}}\ and\ \bibinfo {author} {\bibfnamefont {J.~D.}\ \bibnamefont
  {Dow}},\ }\bibfield  {title} {\bibinfo {title} {{Theory of electroabsorption
  by anisotropic and layered semiconductors. I. Two-dimensional excitons in a
  uniform electric field}},\ }\href {https://doi.org/10.1103/PhysRevB.13.1633}
  {\bibfield  {journal} {\bibinfo  {journal} {Phys. Rev. B}\ }\textbf {\bibinfo
  {volume} {13}},\ \bibinfo {pages} {1633} (\bibinfo {year}
  {1976})}\BibitemShut {NoStop}%
\bibitem [{\citenamefont {Cudazzo}\ \emph {et~al.}(2010)\citenamefont
  {Cudazzo}, \citenamefont {Attaccalite}, \citenamefont {Tokatly},\ and\
  \citenamefont {Rubio}}]{Cudazzo2010}%
  \BibitemOpen
  \bibfield  {author} {\bibinfo {author} {\bibfnamefont {P.}~\bibnamefont
  {Cudazzo}}, \bibinfo {author} {\bibfnamefont {C.}~\bibnamefont
  {Attaccalite}}, \bibinfo {author} {\bibfnamefont {I.~V.}\ \bibnamefont
  {Tokatly}},\ and\ \bibinfo {author} {\bibfnamefont {A.}~\bibnamefont
  {Rubio}},\ }\bibfield  {title} {\bibinfo {title} {{Strong charge-transfer
  excitonic effects and the Bose-Einstein exciton condensate in Graphane}},\
  }\href {https://doi.org/10.1103/PhysRevLett.104.226804} {\bibfield  {journal}
  {\bibinfo  {journal} {Phys. Rev. Lett.}\ }\textbf {\bibinfo {volume} {104}},\
  \bibinfo {pages} {226804} (\bibinfo {year} {2010})}\BibitemShut {NoStop}%
\bibitem [{\citenamefont {Pulci}\ \emph {et~al.}(2012)\citenamefont {Pulci},
  \citenamefont {Gori}, \citenamefont {Marsili}, \citenamefont {Garbuio},
  \citenamefont {{Del Sole}},\ and\ \citenamefont {Bechstedt}}]{Pulci2012a}%
  \BibitemOpen
  \bibfield  {author} {\bibinfo {author} {\bibfnamefont {O.}~\bibnamefont
  {Pulci}}, \bibinfo {author} {\bibfnamefont {P.}~\bibnamefont {Gori}},
  \bibinfo {author} {\bibfnamefont {M.}~\bibnamefont {Marsili}}, \bibinfo
  {author} {\bibfnamefont {V.}~\bibnamefont {Garbuio}}, \bibinfo {author}
  {\bibfnamefont {R.}~\bibnamefont {{Del Sole}}},\ and\ \bibinfo {author}
  {\bibfnamefont {F.}~\bibnamefont {Bechstedt}},\ }\bibfield  {title} {\bibinfo
  {title} {{Strong excitons in novel two-dimensional crystals: Silicane and
  germanane}},\ }\href {https://doi.org/10.1209/0295-5075/98/37004} {\bibfield
  {journal} {\bibinfo  {journal} {EPL}\ }\textbf {\bibinfo {volume} {98}},\
  \bibinfo {pages} {37004} (\bibinfo {year} {2012})}\BibitemShut {NoStop}%
\bibitem [{\citenamefont {Latini}\ \emph {et~al.}(2015)\citenamefont {Latini},
  \citenamefont {Olsen},\ and\ \citenamefont {Thygesen}}]{Latini2015}%
  \BibitemOpen
  \bibfield  {author} {\bibinfo {author} {\bibfnamefont {S.}~\bibnamefont
  {Latini}}, \bibinfo {author} {\bibfnamefont {T.}~\bibnamefont {Olsen}},\ and\
  \bibinfo {author} {\bibfnamefont {K.~S.}\ \bibnamefont {Thygesen}},\
  }\bibfield  {title} {\bibinfo {title} {{Excitons in van der Waals
  heterostructures: The important role of dielectric screening}},\ }\href
  {https://doi.org/10.1103/PhysRevB.92.245123} {\bibfield  {journal} {\bibinfo
  {journal} {Phys. Rev. B}\ }\textbf {\bibinfo {volume} {92}},\ \bibinfo
  {pages} {245123} (\bibinfo {year} {2015})}\BibitemShut {NoStop}%
\bibitem [{\citenamefont {Goryca}\ \emph {et~al.}(2019)\citenamefont {Goryca},
  \citenamefont {Li}, \citenamefont {Stier}, \citenamefont {Taniguchi},
  \citenamefont {Watanabe}, \citenamefont {Courtade}, \citenamefont {Shree},
  \citenamefont {Robert}, \citenamefont {Urbaszek}, \citenamefont {Marie},\
  and\ \citenamefont {Crooker}}]{Goryca2019}%
  \BibitemOpen
  \bibfield  {author} {\bibinfo {author} {\bibfnamefont {M.}~\bibnamefont
  {Goryca}}, \bibinfo {author} {\bibfnamefont {J.}~\bibnamefont {Li}}, \bibinfo
  {author} {\bibfnamefont {A.~V.}\ \bibnamefont {Stier}}, \bibinfo {author}
  {\bibfnamefont {T.}~\bibnamefont {Taniguchi}}, \bibinfo {author}
  {\bibfnamefont {K.}~\bibnamefont {Watanabe}}, \bibinfo {author}
  {\bibfnamefont {E.}~\bibnamefont {Courtade}}, \bibinfo {author}
  {\bibfnamefont {S.}~\bibnamefont {Shree}}, \bibinfo {author} {\bibfnamefont
  {C.}~\bibnamefont {Robert}}, \bibinfo {author} {\bibfnamefont
  {B.}~\bibnamefont {Urbaszek}}, \bibinfo {author} {\bibfnamefont
  {X.}~\bibnamefont {Marie}},\ and\ \bibinfo {author} {\bibfnamefont {S.~A.}\
  \bibnamefont {Crooker}},\ }\bibfield  {title} {\bibinfo {title} {{Revealing
  exciton masses and dielectric properties of monolayer semiconductors with
  high magnetic fields}},\ }\href {https://doi.org/10.1038/s41467-019-12180-y}
  {\bibfield  {journal} {\bibinfo  {journal} {Nat. Commun.}\ }\textbf {\bibinfo
  {volume} {10}},\ \bibinfo {pages} {4172} (\bibinfo {year}
  {2019})}\BibitemShut {NoStop}%
\bibitem [{\citenamefont {Scrinzi}(2010)}]{Scrinzi2010}%
  \BibitemOpen
  \bibfield  {author} {\bibinfo {author} {\bibfnamefont {A.}~\bibnamefont
  {Scrinzi}},\ }\bibfield  {title} {\bibinfo {title} {{Infinite-range exterior
  complex scaling as a perfect absorber in time-dependent problems}},\ }\href
  {https://doi.org/10.1103/PhysRevA.81.053845} {\bibfield  {journal} {\bibinfo
  {journal} {Phys. Rev. A}\ }\textbf {\bibinfo {volume} {81}},\ \bibinfo
  {pages} {053845} (\bibinfo {year} {2010})}\BibitemShut {NoStop}%
\bibitem [{\citenamefont {Kamban}\ \emph
  {et~al.}(2020{\natexlab{b}})\citenamefont {Kamban}, \citenamefont
  {Christensen}, \citenamefont {S{\o}ndergaard},\ and\ \citenamefont
  {Pedersen}}]{Kamban2020}%
  \BibitemOpen
  \bibfield  {author} {\bibinfo {author} {\bibfnamefont {H.~C.}\ \bibnamefont
  {Kamban}}, \bibinfo {author} {\bibfnamefont {S.~S.}\ \bibnamefont
  {Christensen}}, \bibinfo {author} {\bibfnamefont {T.}~\bibnamefont
  {S{\o}ndergaard}},\ and\ \bibinfo {author} {\bibfnamefont {T.~G.}\
  \bibnamefont {Pedersen}},\ }\bibfield  {title} {\bibinfo {title}
  {{Finite-difference time-domain simulation of strong-field ionization: A
  perfectly matched layer approach}},\ }\href
  {https://doi.org/10.1002/pssb.201900467} {\bibfield  {journal} {\bibinfo
  {journal} {Phys. Stat. Sol. B}\ }\textbf {\bibinfo {volume} {257}},\ \bibinfo
  {pages} {1900467} (\bibinfo {year} {2020}{\natexlab{b}})}\BibitemShut
  {NoStop}%
\bibitem [{\citenamefont {Langer}\ \emph {et~al.}(2018)\citenamefont {Langer},
  \citenamefont {Schmid}, \citenamefont {Schlauderer}, \citenamefont {Gmitra},
  \citenamefont {Fabian}, \citenamefont {Nagler}, \citenamefont
  {Sch{\"{u}}ller}, \citenamefont {Korn}, \citenamefont {Hawkins},
  \citenamefont {Steiner}, \citenamefont {Huttner}, \citenamefont {Koch},
  \citenamefont {Kira},\ and\ \citenamefont {Huber}}]{Langer2018}%
  \BibitemOpen
  \bibfield  {author} {\bibinfo {author} {\bibfnamefont {F.}~\bibnamefont
  {Langer}}, \bibinfo {author} {\bibfnamefont {C.~P.}\ \bibnamefont {Schmid}},
  \bibinfo {author} {\bibfnamefont {S.}~\bibnamefont {Schlauderer}}, \bibinfo
  {author} {\bibfnamefont {M.}~\bibnamefont {Gmitra}}, \bibinfo {author}
  {\bibfnamefont {J.}~\bibnamefont {Fabian}}, \bibinfo {author} {\bibfnamefont
  {P.}~\bibnamefont {Nagler}}, \bibinfo {author} {\bibfnamefont
  {C.}~\bibnamefont {Sch{\"{u}}ller}}, \bibinfo {author} {\bibfnamefont
  {T.}~\bibnamefont {Korn}}, \bibinfo {author} {\bibfnamefont {P.~G.}\
  \bibnamefont {Hawkins}}, \bibinfo {author} {\bibfnamefont {J.~T.}\
  \bibnamefont {Steiner}}, \bibinfo {author} {\bibfnamefont {U.}~\bibnamefont
  {Huttner}}, \bibinfo {author} {\bibfnamefont {S.~W.}\ \bibnamefont {Koch}},
  \bibinfo {author} {\bibfnamefont {M.}~\bibnamefont {Kira}},\ and\ \bibinfo
  {author} {\bibfnamefont {R.}~\bibnamefont {Huber}},\ }\bibfield  {title}
  {\bibinfo {title} {{Lightwave valleytronics in a monolayer of tungsten
  diselenide}},\ }\href {https://doi.org/10.1038/s41586-018-0013-6} {\bibfield
  {journal} {\bibinfo  {journal} {Nature}\ }\textbf {\bibinfo {volume} {557}},\
  \bibinfo {pages} {76} (\bibinfo {year} {2018})}\BibitemShut {NoStop}%
\bibitem [{\citenamefont {Yang}\ and\ \citenamefont
  {Robicheaux}(2014)}]{Yang2014}%
  \BibitemOpen
  \bibfield  {author} {\bibinfo {author} {\bibfnamefont {B.~C.}\ \bibnamefont
  {Yang}}\ and\ \bibinfo {author} {\bibfnamefont {F.}~\bibnamefont
  {Robicheaux}},\ }\bibfield  {title} {\bibinfo {title} {{Field-ionization
  threshold and its induced ionization-window phenomenon for Rydberg atoms in a
  short single-cycle pulse}},\ }\href
  {https://doi.org/10.1103/PhysRevA.90.063413} {\bibfield  {journal} {\bibinfo
  {journal} {Phys. Rev. A}\ }\textbf {\bibinfo {volume} {90}},\ \bibinfo
  {pages} {063413} (\bibinfo {year} {2014})}\BibitemShut {NoStop}%
\bibitem [{\citenamefont {Henriques}\ \emph {et~al.}(2020)\citenamefont
  {Henriques}, \citenamefont {Kamban}, \citenamefont {Pedersen},\ and\
  \citenamefont {Peres}}]{Henriques2020a}%
  \BibitemOpen
  \bibfield  {author} {\bibinfo {author} {\bibfnamefont {J.~C.}\ \bibnamefont
  {Henriques}}, \bibinfo {author} {\bibfnamefont {H.~C.}\ \bibnamefont
  {Kamban}}, \bibinfo {author} {\bibfnamefont {T.~G.}\ \bibnamefont
  {Pedersen}},\ and\ \bibinfo {author} {\bibfnamefont {N.~M.}\ \bibnamefont
  {Peres}},\ }\bibfield  {title} {\bibinfo {title} {{Analytical quantitative
  semiclassical approach to the Lo Surdo-Stark effect and ionization in
  two-dimensional excitons}},\ }\href
  {https://doi.org/10.1103/PhysRevB.102.035402} {\bibfield  {journal} {\bibinfo
   {journal} {Phys. Rev. B}\ }\textbf {\bibinfo {volume} {102}},\ \bibinfo
  {pages} {035402} (\bibinfo {year} {2020})}\BibitemShut {NoStop}%
\bibitem [{\citenamefont {Abramowitz}\ and\ \citenamefont
  {Stegun}(1972)}]{Abramowitz1972}%
  \BibitemOpen
  \bibinfo {editor} {\bibfnamefont {M.}~\bibnamefont {Abramowitz}}\ and\
  \bibinfo {editor} {\bibfnamefont {I.}~\bibnamefont {Stegun}},\ eds.,\
  \href@noop {} {\emph {\bibinfo {title} {{Handbook of Mathematical Functions,
  With Formulas, Graphs, and Mathematical Tables}}}}\ (\bibinfo  {publisher}
  {Dover},\ \bibinfo {address} {New York},\ \bibinfo {year} {1972})\BibitemShut
  {NoStop}%
\bibitem [{\citenamefont {{William H. Press}}\ \emph
  {et~al.}(2007)\citenamefont {{William H. Press}}, \citenamefont {Teukolsky},
  \citenamefont {Vetterling},\ and\ \citenamefont {Flannery}}]{William2007}%
  \BibitemOpen
  \bibfield  {author} {\bibinfo {author} {\bibnamefont {{William H. Press}}},
  \bibinfo {author} {\bibfnamefont {S.~A.}\ \bibnamefont {Teukolsky}}, \bibinfo
  {author} {\bibfnamefont {W.~T.}\ \bibnamefont {Vetterling}},\ and\ \bibinfo
  {author} {\bibfnamefont {B.~P.}\ \bibnamefont {Flannery}},\ }\href@noop {}
  {\emph {\bibinfo {title} {{Numerical Recipes}}}},\ \bibinfo {edition} {3rd}\
  ed.\ (\bibinfo  {publisher} {Cambridge University Press},\ \bibinfo {address}
  {New York},\ \bibinfo {year} {2007})\BibitemShut {NoStop}%
\bibitem [{\citenamefont {Bengtsson}\ \emph {et~al.}(2008)\citenamefont
  {Bengtsson}, \citenamefont {Lindroth},\ and\ \citenamefont
  {Selst{\o}}}]{Bengtsson2008}%
  \BibitemOpen
  \bibfield  {author} {\bibinfo {author} {\bibfnamefont {J.}~\bibnamefont
  {Bengtsson}}, \bibinfo {author} {\bibfnamefont {E.}~\bibnamefont
  {Lindroth}},\ and\ \bibinfo {author} {\bibfnamefont {S.}~\bibnamefont
  {Selst{\o}}},\ }\bibfield  {title} {\bibinfo {title} {{Solution of the
  time-dependent Schr{\"{o}}dinger equation using uniform complex scaling}},\
  }\href {https://doi.org/10.1103/PhysRevA.78.032502} {\bibfield  {journal}
  {\bibinfo  {journal} {Phys. Rev. A}\ }\textbf {\bibinfo {volume} {78}},\
  \bibinfo {pages} {032502} (\bibinfo {year} {2008})}\BibitemShut {NoStop}%
\bibitem [{\citenamefont {Perea-Caus{\'{i}}n}\ \emph
  {et~al.}(2021)\citenamefont {Perea-Caus{\'{i}}n}, \citenamefont {Brem},\ and\
  \citenamefont {Malic}}]{Perea-Causin2021}%
  \BibitemOpen
  \bibfield  {author} {\bibinfo {author} {\bibfnamefont {R.}~\bibnamefont
  {Perea-Caus{\'{i}}n}}, \bibinfo {author} {\bibfnamefont {S.}~\bibnamefont
  {Brem}},\ and\ \bibinfo {author} {\bibfnamefont {E.}~\bibnamefont {Malic}},\
  }\bibfield  {title} {\bibinfo {title} {{Phonon-assisted exciton dissociation
  in transition metal dichalcogenides}},\ }\href
  {https://doi.org/10.1039/d0nr07131e} {\bibfield  {journal} {\bibinfo
  {journal} {Nanoscale}\ }\textbf {\bibinfo {volume} {13}},\ \bibinfo {pages}
  {1884} (\bibinfo {year} {2021})}\BibitemShut {NoStop}%
\bibitem [{\citenamefont {Haastrup}\ \emph {et~al.}(2018)\citenamefont
  {Haastrup}, \citenamefont {Strange}, \citenamefont {Pandey}, \citenamefont
  {Deilmann}, \citenamefont {Schmidt}, \citenamefont {Hinsche}, \citenamefont
  {Gjerding}, \citenamefont {Torelli}, \citenamefont {Larsen}, \citenamefont
  {Riis-Jensen}, \citenamefont {Gath}, \citenamefont {Jacobsen}, \citenamefont
  {Mortensen}, \citenamefont {Olsen},\ and\ \citenamefont
  {Thygesen}}]{Haastrup2018}%
  \BibitemOpen
  \bibfield  {author} {\bibinfo {author} {\bibfnamefont {S.}~\bibnamefont
  {Haastrup}}, \bibinfo {author} {\bibfnamefont {M.}~\bibnamefont {Strange}},
  \bibinfo {author} {\bibfnamefont {M.}~\bibnamefont {Pandey}}, \bibinfo
  {author} {\bibfnamefont {T.}~\bibnamefont {Deilmann}}, \bibinfo {author}
  {\bibfnamefont {P.~S.}\ \bibnamefont {Schmidt}}, \bibinfo {author}
  {\bibfnamefont {N.~F.}\ \bibnamefont {Hinsche}}, \bibinfo {author}
  {\bibfnamefont {M.~N.}\ \bibnamefont {Gjerding}}, \bibinfo {author}
  {\bibfnamefont {D.}~\bibnamefont {Torelli}}, \bibinfo {author} {\bibfnamefont
  {P.~M.}\ \bibnamefont {Larsen}}, \bibinfo {author} {\bibfnamefont {A.~C.}\
  \bibnamefont {Riis-Jensen}}, \bibinfo {author} {\bibfnamefont
  {J.}~\bibnamefont {Gath}}, \bibinfo {author} {\bibfnamefont {K.~W.}\
  \bibnamefont {Jacobsen}}, \bibinfo {author} {\bibfnamefont {J.~J.}\
  \bibnamefont {Mortensen}}, \bibinfo {author} {\bibfnamefont {T.}~\bibnamefont
  {Olsen}},\ and\ \bibinfo {author} {\bibfnamefont {K.~S.}\ \bibnamefont
  {Thygesen}},\ }\bibfield  {title} {\bibinfo {title} {{The Computational 2D
  Materials Database: High-throughput modeling and discovery of atomically thin
  crystals}},\ }\href {https://doi.org/10.1088/2053-1583/aacfc1} {\bibfield
  {journal} {\bibinfo  {journal} {2D Mater.}\ }\textbf {\bibinfo {volume}
  {5}},\ \bibinfo {pages} {042002} (\bibinfo {year} {2018})}\BibitemShut
  {NoStop}%
\bibitem [{\citenamefont {Dargys}\ and\ \citenamefont
  {Kundrotas}(1998)}]{Dargys1998}%
  \BibitemOpen
  \bibfield  {author} {\bibinfo {author} {\bibfnamefont {A.}~\bibnamefont
  {Dargys}}\ and\ \bibinfo {author} {\bibfnamefont {J.}~\bibnamefont
  {Kundrotas}},\ }\bibfield  {title} {\bibinfo {title} {{Impact ionization of
  excitons by hot carriers in quantum wells}},\ }\href
  {https://doi.org/10.1088/0268-1242/13/11/004} {\bibfield  {journal} {\bibinfo
   {journal} {Semicond. Sci. Technol.}\ }\textbf {\bibinfo {volume} {13}},\
  \bibinfo {pages} {1258} (\bibinfo {year} {1998})}\BibitemShut {NoStop}%
\bibitem [{\citenamefont {Kaindl}\ \emph {et~al.}(2009)\citenamefont {Kaindl},
  \citenamefont {H{\"{a}}gele}, \citenamefont {Carnahan},\ and\ \citenamefont
  {Chemla}}]{Kaindl2009}%
  \BibitemOpen
  \bibfield  {author} {\bibinfo {author} {\bibfnamefont {R.~A.}\ \bibnamefont
  {Kaindl}}, \bibinfo {author} {\bibfnamefont {D.}~\bibnamefont
  {H{\"{a}}gele}}, \bibinfo {author} {\bibfnamefont {M.~A.}\ \bibnamefont
  {Carnahan}},\ and\ \bibinfo {author} {\bibfnamefont {D.~S.}\ \bibnamefont
  {Chemla}},\ }\bibfield  {title} {\bibinfo {title} {{Transient terahertz
  spectroscopy of excitons and unbound carriers in quasi-two-dimensional
  electron-hole gases}},\ }\href {https://doi.org/10.1103/PhysRevB.79.045320}
  {\bibfield  {journal} {\bibinfo  {journal} {Phys. Rev. B}\ }\textbf {\bibinfo
  {volume} {79}},\ \bibinfo {pages} {045320} (\bibinfo {year}
  {2009})}\BibitemShut {NoStop}%
\bibitem [{\citenamefont {Aivazian}\ \emph {et~al.}(2017)\citenamefont
  {Aivazian}, \citenamefont {Yu}, \citenamefont {Wu}, \citenamefont {Yan},
  \citenamefont {Mandrus}, \citenamefont {Cobden}, \citenamefont {Yao},\ and\
  \citenamefont {Xu}}]{Aivazian2017}%
  \BibitemOpen
  \bibfield  {author} {\bibinfo {author} {\bibfnamefont {G.}~\bibnamefont
  {Aivazian}}, \bibinfo {author} {\bibfnamefont {H.}~\bibnamefont {Yu}},
  \bibinfo {author} {\bibfnamefont {S.}~\bibnamefont {Wu}}, \bibinfo {author}
  {\bibfnamefont {J.}~\bibnamefont {Yan}}, \bibinfo {author} {\bibfnamefont
  {D.~G.}\ \bibnamefont {Mandrus}}, \bibinfo {author} {\bibfnamefont
  {D.}~\bibnamefont {Cobden}}, \bibinfo {author} {\bibfnamefont
  {W.}~\bibnamefont {Yao}},\ and\ \bibinfo {author} {\bibfnamefont
  {X.}~\bibnamefont {Xu}},\ }\bibfield  {title} {\bibinfo {title} {{Many-body
  effects in nonlinear optical responses of 2D layered semiconductors}},\
  }\href {https://doi.org/10.1088/2053-1583/aa56f1} {\bibfield  {journal}
  {\bibinfo  {journal} {2D Mater.}\ }\textbf {\bibinfo {volume} {4}},\ \bibinfo
  {pages} {025024} (\bibinfo {year} {2017})}\BibitemShut {NoStop}%
\end{thebibliography}%

\end{document}